\documentclass{ptephy_v1}

\preprintnumber{XXXX-XXXX} 

\newcommand{\delslash}{\partial \hspace{-6pt}/ \hspace{1pt}}
\usepackage{bm}

\usepackage[normalem]{ulem} 
\usepackage{color}

\renewcommand\sout{\bgroup \color{black} \ULdepth=-.5ex \ULset}

\begin{document}

\title{Dynamical supersymmetry for strange quark and $ud$ antidiquark in hadron mass spectrum}


\author[1,2]{Taiju Amano}
\author[2,1]{Daisuke Jido}
\affil[1]{Department of Physics, Tokyo Metropolitan University, 1-1 Minami-Osawa, Hachioji, 192-0397 Tokyo, Japan \email{amano-taiju@ed.tmu.ac.jp}}
\affil[2]{Department of Physics, Tokyo Institute of Technology, 2-12-1 Ookayama, Meguro-ku, Tokyo 152-8550 Japan}


\begin{abstract}%
Speculating that the $ud$ diquark with spin 0 has a similar mass to the constituent $s$ quark, we introduce a symmetry between the $s$ quark and the $\overline{ud}$ diquark. Constructing an algebra for this symmetry, we regard a triplet of the $s$ quarks with spin up and down and the $\overline{ud}$ diquark with spin 0 as a fundamental representation of this algebra. We further build higher representations constructed by direct products of the fundamental representations. We propose assignments of hadrons to the multiples of this algebra.
We find in particular that $\{D_{s}, D_{s}^{*}, \Lambda_{c}\}$, $\{\eta_{s}, \phi, \Lambda, f_{0}(1370)\}$ and $\{\Omega_{c}, T_{sc}\}$ form a triplet, a nonet and a quintet, respectively, where $T_{sc}$ is a genuine tetraquark meson composed of $\overline{ud}sc$.
We also find a mass relation between them by introducing the symmetry breaking due to the mass difference between the $s$ quark and the $\overline{ud}$ diquark. 
The masses of possible tetraquarks $\overline{ud}sc$ and $\overline{ud}sb$ are estimated from
the symmetry breaking and the masses of $\Omega_{c}$ and $\Omega_{b}$ to be 2.942~GeV and 6.261~GeV, 
respectively.
\end{abstract}

\subjectindex{xxxx, xxx}
\maketitle

\section{Introduction}
Symmetries play important roles in hadron physics. Hadrons can be classified into the representations of symmetry groups, and the hadron masses and interactions can be explained by the symmetry properties. In particular, the flavor SU(3) symmetry is one of the most successful examples to understand the hadron spectra. After having discovered strangeness, one collects hadrons having a similar mass and classifies into octets and decuplet of the SU(3) representation~\cite{Neeman:1961jhl,GellMann:1962xb} according to the celebrated Gell-Mann Nishijima relation~\cite{Nakano:1953zz,GellMann:1953zza}. Behind this classification, the up, down and strange quarks are found as the fundamental representation of the symmetry~\cite{GellMann:1964nj}. The flavor SU(3) symmetry is not exact but is broken explicitly with the quark mass difference. The symmetry breaking pattern is also constrained by the symmetry properties. Treating the quark mass difference as a first order perturbation, one obtains the so-called Gell-Mann Okubo mass formulae~\cite{Okubo:1961jc,GellMann:1962xb} which relates the masses of the hadrons in the same multiplet. In this way, one finds the substantial objects which carry the fundamental properties of symmetry out of the hadron spectrum. In this paper, regarding the constituent strange quark and the $\overline{ud}$ diquark as a fundamental object of a symmetry, we find mass relations among the hadrons classified in the same multiplet of the symmetry and discuss the possibility of the existence of the $ud$ diquark as an effective constituent of hadrons.

The diquark is a pair of two quarks and is considered as a strong candidate of the hadron consituent~\cite{Ida:1966ev,Lichtenberg:1967zz,Anselmino:1992vg}. Because the diquark has color charge, it cannot be isolated and should exist inside hadrons. One expects a strong correlation particularly between up and down quarks with spin 0 and isospin 0 due to color magnetic interaction~\cite{Jaffe:2004ph} and such strong correlations are also found in Lattice QCD studies~\cite{Hess98,Babich,Orginos,Alexandrou}. The diquark has been investigated in a context of the quark models in Refs.~\cite{Goldstein:1979wba,Lichtenberg:1981pp,Lichtenberg:1982jp,Liu:1983us,Ebert:2005xj,Ebert:2007nw,Hernandez:2008ej,Lee:2009rt}, and recently it has been found in Refs.~\cite{Jido:2016yuv,Kumakawa:2017ffl} that the color electric force between diquark and quark could be weaker than that between quark and antiquark. A QCD sum rule approach~\cite{Kim:2011ut} has suggested also the $ud$ diquark as a constituent of the ground states of $\Lambda$, $\Lambda_{c}$ and $\Lambda_{b}$ having a constituent diquark mass around 0.4 GeV. The mass of the $ud$ diquark is not fixed yet. Considering the $u$ and $d$ constituent quark mass to be about 0.3 GeV, one expects that the diquark mass be 0.4 to 0.6 GeV depending on the attraction between the $u$ and $d$ quark. Such value of the diquark mass is very similar to the constituent strange mass, which may be 0.5 GeV.

In this paper, we introduce a symmetry in which the constituent $s$ quark and the $\overline{ud}$ diquark form a fundamental representation thanks to their similar masses and classify hadrons according to the symmetry to discuss the breaking pattern of the symmetry in the mass spectrum of hadrons composed of the $s$ quark and $\overline{ud}$ diquark. This is the same approach to find the flavor SU(3) symmetry in the hadron spectrum. While both the $s$ quark  and the $\overline{ud}$ diquark have the same color charge, they have different spins; the $s$ quark is a fermion with spin 1/2 and the $\overline{ud}$ diquark is a boson with spin 0. Thus, the symmetry that we consider here is a supersymmetry which transforms fermions and bosons. 
This symmetry is not a fundamental symmetry of QCD, but a symmetry among objects as outcomes of QCD dynamics. The elements of the symmetry considered here are the strange constituent quark and the $ud$ antidiquark. They are quasiparticles obtained by quark-gluon dynamics and can be regarded as effective constituents of the hadron structure. Thus, we examine the possibility to have such kinds of symmetry in the hadron spectra. Here we consider the static properties of hadrons by assuming that the hadrons are composed of these effective constituents. For the grand state hadrons, there may be enough time to form the constituent quark and diquark inside the hadrons.
The supersymmetry was introduced first in hadron physics in Refs.~\cite{miya66,miya68}. There ($p$,$n$,$\Lambda$) and ($K^{+}$,$K^{0}$,$\eta$) were considered as flavor fundamental representations, and a supersymmetry between these two triplets were investigated using so-called V(3) algebra. A supersymmetry between quark and diquark was discussed also in Refs.~\cite{Catto:1984wi,Lichtenberg:1989ix,Nielsen:2018uyn}. Dynamical supersymmetry in nuclear physics was suggested in Ref.~\cite{Iachello:1980av}.

In this paper, in Sec.~2 we define the algebra in which the spin up and down $s$ quark and the $\overline{ud}$ diquark with spin 0 form a triplet. In Sec.~3, we discuss the representation of the algebra introduced in Sec.~2, and show examples of the representations for hadrons in Sec.~4. Section~5 presents the symmetry breaking by the mass difference of the $s$ quark and the $\overline{ud}$ diquark, and derives a Gell-Mann Okubo type mass formula for $\phi$, $\Lambda$ and $f_{0}$. Section~6 is devoted to summary and conclusion. 

\section{Definition of algebra}

In this section, we introduce a supersymmetry for a Dirac fermion 
with spin $1/2$ and a charged scalar boson in the flavor space
according to Ref.~\cite{miya66}.

\subsection{Field definition}

Let us first define the fermion and boson fields. 
We write the fermion and boson fields as $\psi$ and $\varphi$, respectively.
The fermion field $\psi$ has four components, two of them are so-called
upper components in the Dirac representation, the others are 
the lower components, while the scalar field $\varphi$ is composed 
of two independent real fields for a charged boson. 

The Lagrangians for the free Dirac field and the scalar boson field are
written as
\begin{align}
   {\cal L}_{F} &= \bar \psi i\delslash \psi - m \bar \psi \psi \\
   {\cal L}_{B} &= \partial_{\mu} \varphi^{\dagger} \partial^{\mu} \varphi
  - m^{2} \varphi^{\dagger} \varphi,
\end{align}
respectively. Defining the conjugate momenta as
\begin{equation}
   \bar \pi = \frac{\partial {\cal L}_{F}}{\partial \dot\psi} = i \psi^{\dagger}, \qquad
   \pi = \frac{\partial {\cal L}_{B}}{\partial \dot \varphi} = \dot \varphi^{\dagger}, \qquad
   \pi^{\dagger} = \frac{\partial {\cal L}_{B}}{\partial \dot \varphi^{\dagger}} = \dot \varphi,
\end{equation}
we have the corresponding Hamiltonians as
\begin{align}
   {\cal H}_{F} &= \bar \pi \dot \psi - {\cal L}_{F} 
   = \bar \psi i \vec \nabla \cdot \vec \gamma \psi + m \bar \psi \psi \\
   {\cal H}_{B} &= \pi \dot \varphi + \dot \varphi^{\dagger} \pi^{\dagger} - {\cal L}_{B}
   = \vec \nabla \varphi^{\dagger} \cdot \vec \nabla \varphi 
   + m (\frac{1}{m} \pi^{\dagger} \pi + m \varphi^{\dagger} \varphi).
\end{align}

Quantization is performed by introducing the equal-time commutation 
relations for the fermion and boson fields.
The field commutation relations are given as
\begin{equation}
    \{ \psi_\alpha(x), \psi_\beta^\dag(y) \} = \delta_{\alpha \beta}  \delta(x-y),
\end{equation}
for fermion, where $\alpha$ and $\beta$ stand for the Dirac components,
and 
\begin{equation}
    [ \varphi(x), \pi(y) ] = \delta(x-y), \qquad
    [ \varphi^{\dagger}(x), \pi^{\dagger}(y) ] = \delta(x-y),
\end{equation}
for boson. These expressions are not symmetric in terms of the fermion 
and boson fields. In the following we redefine the fields in a symmetric 
form.

Let us introduce two-component fields, $\psi^{(+)}$ and $\psi^{(-)}$, as
the eigenfunction of $\gamma_{0}$ with eigenvalue $\pm 1$, respectively.
In the Dirac representation, $\psi^{(+)}$ and $\psi^{(-)}$ are 
the upper and lower components of the Dirac field, respectively, as
\begin{equation}
   \psi = \left(\begin{array}{c} \psi^{(+)} \\ \psi^{(-)} \end{array} \right).
\end{equation}
Their conjugate fields are denoted by
\begin{equation}
  \hat \psi^{(\pm)} \equiv \bar \psi^{(\pm)} \gamma_{0} = \psi^{(\pm)\dagger} \label{eq:conjugate fermion}
\end{equation}
Using the $\hat \psi^{(\pm)}$ field, the anti-commutation relation of 
the fermion field is written as
\begin{equation}
   \{ \psi_i^{(+)}(x), \hat{\psi}_j^{(+)}(y) \}
   = \{ \psi_{i}^{(-)}(x), \hat{\psi}_{j}^{(-)}(y) \} 
   = \delta_{ij} \delta(x-y) \label{eq:fcom}
\end{equation}
for $i,j = 1,2$, and the $\psi^{(+)}$ and $\psi^{(-)}$ are anti-commuting.

The pseudoscalar and vector feilds, $\bar \psi \gamma_{5} \psi$ and $\bar \psi \gamma^{i} \psi$, 
are decomposed into $\hat \psi^{(\pm)}$ and $\psi^{(\pm)}$ as
\begin{align}
   \bar \psi \gamma_{5} \psi &= - \hat \psi^{(-)} \psi^{(+)}  + \hat \psi^{(+)} \psi^{(-)}, \\
   \bar \psi \gamma^{i} \psi &=  \hat \psi^{(-)} \sigma_{i} \psi^{(+)} + \hat \psi^{(+)} \sigma_{i} \psi^{(-)},
\end{align}
where $\sigma_{i}$ is the Pauli matrix in the spin space. 

For the boson field, we introduce the following two independent fields
\begin{equation}
  \begin{split}
  \varphi^{(+)} &= \sqrt\frac m2 \varphi + i \sqrt\frac1{2m} \pi^{\dagger} \\
  \varphi^{(-)} &= \sqrt\frac m2 \varphi - i \sqrt\frac1{2m} \pi^{\dagger} 
\end{split}
\end{equation}
and their conjugate fields are denoted by
\begin{equation}
\begin{split}
 \bar \varphi^{(+)} &= (\varphi^{(+)})^{\dagger}
 = \sqrt\frac m2 \varphi^{\dagger} - i \sqrt\frac1{2m} \pi  \\
  \bar \varphi^{(-)} &  = (\varphi^{(-)})^{\dagger}
= \sqrt\frac m2 \varphi^{\dagger} + i \sqrt\frac1{2m} \pi 
  \end{split}
\end{equation}
Now, in the similar way to the fermion field, we introduce 
\begin{equation}
   \hat \varphi^{(\pm)} \equiv \bar \varphi^{(\pm)} \gamma_{0} = \pm (\varphi^{(\pm)})^{\dagger} \label{eq:conjugate boson}
\end{equation}
where $\gamma_{0}$ for the boson field is a 2 by 2 matrix and 
$\varphi^{(\pm)}$ is the eigenvector of $\gamma_{0}$ with eigenvalue
$\pm1$. It is easy to check that the boson fields $\varphi^{(\pm)}$
and $\hat \varphi^{(\pm)}$ satisfy the following commutation relation:
\begin{equation}
  [ \varphi^{(+)}(x), \hat{\varphi}^{(+)}(y) ] 
  = [ \varphi^{(-)}(x), \hat{\varphi}^{(-)}(y) ] =  \delta(x-y) \label{eq:bcom}
\end{equation}
The $\varphi^{(\pm)}$ and $\hat \varphi^{(\pm)}$ fields are commuting. 
Now we have the commutation relations \eqref{eq:fcom} and \eqref{eq:bcom}
in a symmetric form. 

Writing the mass term of the Hamiltonians in the redefined fields,
we obtain 
\begin{equation}
   {\cal H}_{\rm mass} = m (\hat \psi^{(+)}_{1} \psi^{(+)}_{1} + \hat \psi^{(+)}_{2} \psi^{(+)}_{2}
   + \hat \varphi^{(+)} \varphi^{(+)}
    - \hat \psi^{(-)}_{1} \psi^{(-)}_{1}  - \hat \psi^{(-)}_{2} \psi^{(-)}_{2}
    - \hat \varphi^{(-)} \varphi^{(-)}), \label{eq:Hmass}
\end{equation}
if one assumes the same mass $m$ for the fermion and boson.

\subsection{V(3) algebra}
\subsubsection{Generators of V(3)}
Now let us consider the fermion and boson fields as a triplet for each 
$(\pm)$ component:
\begin{equation}
   \Psi^{(\pm)} = \left(\begin{array}{c} \psi_{1}^{(\pm)} \\ \psi_{2}^{(\pm)} 
   \\ \varphi^{(\pm)} \end{array} \right).
\end{equation}
Hereafter we indicate the $(+)$ and $(-)$ components by 
the superscript and subscript, respectively:
\begin{equation}
   \Psi^{(+)}_{i}  = \Psi^{i} =  \left(\begin{array}{c} \psi^{1} \\ \psi^{2} 
   \\ \varphi^{3} \end{array} \right), \qquad
   \Psi^{(-)}_{i}  = \Psi_{i} =  \left(\begin{array}{c} \psi_{1} \\ \psi_{2} 
   \\ \varphi_{3} \end{array} \right) \label{eq: pm field}
\end{equation}
We introduce transformation among the triplet for each component. 
We call this algebra by V(3) accordingly to Ref.~\cite{miya68}. 
This algebra has SU(2) as a subalgebra.
The fermion field is transformed as a doublet of SU(2), 
while the boson field is transformed as a singlet of SU(2). 
Regarding the fermion field as quark and the boson field as antidiquark,
we also introduce baryon number. The fermion field has baryon number $1/3$,
while the boson field has baryon number $-2/3$. We can label each component
of the V(3) representation by the 3rd component of spin and the baryon number
$(S_{3}, B)$ in the similar way of the isospin 3rd component and hypercharge 
$(I_{3}, Y)$ for SU(3). 

The generators
of these transformations can be written as
\begin{align}
   G^{ij} &= \int \hat \psi^{i}(x) \psi^{j}(x) d^{3}x, \qquad 
   G^{33} = \int \hat \varphi^{3}(x) \varphi^{3}(x) d^{3}x, \\
   G_{ij} &= \int \hat \psi_{i}(x) \psi_{j}(x) d^{3}x, \qquad 
   G_{33} = \int \hat \varphi_{3}(x) \varphi_{3}(x) d^{3}x,
\end{align}
for $i,j=1,2$.
With these generators, the triplets transform as
\begin{align}
 & [G^{ij}, \psi^{k}] = - \delta_{ik} \psi^{j}, \qquad
  [G^{ij}, \hat \psi^{k}] = \delta_{jk} \hat \psi^{i}, \qquad
  [G^{33},  \varphi^{3}] = - \varphi^{3}, \qquad
  [G^{33}, \hat \varphi^{3}] = \hat \varphi^{3} \\
 & [G_{ij}, \psi_{k}] = - \delta_{ik} \psi_{j}, \qquad
  [G_{ij}, \hat \psi_{k}] = \delta_{jk} \hat \psi_{i}, \qquad
  [G_{33},  \varphi_{3}] = - \varphi_{3}, \qquad
  [G_{33}, \hat \varphi_{3}] = \hat \varphi_{3},
\end{align}
and 
\begin{align}
  &[G^{ij},\varphi^{3}] = [G^{ij}, \hat \varphi^{3}] 
  = [G^{33}, \psi^{i}] = [G^{33}, \hat \psi^{i}] =0, \\
  & [G_{ij},\varphi_{3}] = [G_{ij}, \hat \varphi_{3}] 
  = [G_{33}, \psi_{i}] = [G_{33}, \hat \psi_{i}]  
  =0,
\end{align}
for $i,j=1,2$. These transformation rules is easily checked 
by using the commutation relations for the fields \eqref{eq:fcom} and \eqref{eq:bcom}.

We also introduce the generators
which transform fermion and boson as
\begin{align}
   G^{i3} &= \int \hat \psi^{i}(x) \varphi^{3}(x) d^{3}x, \qquad
   G^{3i} = \int \hat \varphi^{3}(x) \psi^{i}(x) d^{3}x, \\
   G_{i3} &= \int \hat \psi_{i}(x) \varphi_{3}(x) d^{3}x, \qquad
   G_{3i} = \int \hat \varphi_{3}(x) \psi_{i}(x) d^{3}x, 
\end{align}
for $i=1,2$. These generators interchange the fermion and boson 
fields. For instance, we have
\begin{align}
   \{ G^{i3}, \psi^{j} \} &= \int \{ \hat \psi^{i} \varphi^{3}, \psi^{j}\} d^{3}x
   =  \int \{ \hat \psi^{i},  \psi^{j} \} \varphi^{3} d^{3}x
   - \int \hat \psi^{i} [ \varphi^{3}, \psi^{j} ] d^{3}x = \delta_{ij} \varphi^{3},
\end{align}
where we have used the commutation relation \eqref{eq:fcom}.
Here it should be noted that the generator $G^{i3}$ and field $\psi^{j}$
are fermionic and one should use the anticommutation relation
for the transformation. 
The other transformation rules are
\begin{align}  
  [G^{i3}, \hat \varphi^{3}] &= \hat \psi^{i}, \qquad 
  \{G^{3i}, \hat \psi^{j} \} = \delta_{ij} \hat\varphi^{3}, \qquad
  [G^{3i}, \varphi^{3} ] =  - \psi^{i}, \\
  \{G_{i3}, \psi_{j}\} = \delta_{ij} \varphi_{3}, \qquad
  [G_{i3}, \hat \varphi_{3}] &= \hat \psi_{i}, \qquad 
  \{G_{3i}, \hat \psi_{j} \} = \delta_{ij} \hat \varphi_{3}, \qquad
  [G_{3i}, \varphi_{3} ] =  - \psi_{i}, 
\end{align}
and 
\begin{align}
  \{G^{i3}, \hat \psi^{j}\} &= [G^{i3},  \varphi^{3}] =
  \{G^{3i}, \psi^{j} \} = [G^{3i}, \hat \varphi^{3} ] = 0, \\
  \{G_{i3}, \hat \psi_{j}\} &=  [G_{i3},  \varphi_{3}] =
  \{G_{3i}, \psi_{j} \} =  [G_{3i}, \hat  \varphi_{3} ] =  0. 
\end{align}

\subsubsection{Commutation relations of generators}

Now let us show the commutation relations for the V(3) generators.
The indices $i,j,k,l$ stand for $1,2$. 
The commutation relations for the $1,2$ components read
\begin{equation}
   [G^{ij}, G^{kl}] = \delta_{jk} G^{il} - \delta_{il} G^{kj}, \qquad
   [G_{ij}, G_{kl}] = \delta_{jk} G_{il} - \delta_{il} G_{kj}.
\end{equation}
Some of these commutation relations can be written as
\begin{equation}
\  [G^{a}, G^{b}] = i \epsilon^{abc} G^{c},  \qquad
   [G_{a}, G_{b}] = i \epsilon^{abc} G_{c}, 
\end{equation}
where we have defined 
\begin{equation}
    G^{a} = \int \hat \psi^{i}(x) \frac{(\lambda_{a})_{ij}}2\psi^{j}(x) d^{3}x, \qquad 
    G_{a} = \int \hat \psi_{i}(x) \frac{(\lambda_{a})_{ij}}2\psi_{j}(x) d^{3}x
\end{equation}
with the Gell-man matrix $\lambda_{a}$ for $a=1,2,3$. There are the generators of the SU(2) 
subalgebra. 
We also introduce the operator for the baryon number as
\begin{equation}
\begin{split}
   G^{8} = \frac{1}{3} G^{11} + \frac{1}{3} G^{22} - \frac23 G^{33} 
   = \frac{1}{\sqrt 3} \int \hat \psi^{i}(x) (\lambda_{8})_{ij}\psi^{j}(x) d^{3}x, \\
   G_{8} = \frac{1}{3} G_{11} + \frac{1}{3} G_{22} - \frac23 G_{33} 
   = \frac{1}{\sqrt 3} \int \hat \psi_{i}(x) (\lambda_{8})_{ij}\psi_{j}(x) d^{3}x,
\end{split}
\end{equation}
with $\lambda_{8}$ being the eighth component of the Gell-Man matrix.  
The other commutation relations for the bosonic generators
vanish:
\begin{equation}
   [G^{ij},G^{33}]=[G^{33}, G^{33}] = 0, \qquad
   [G_{ij},G_{33}]=[G_{33}, G_{33}] = 0.
\end{equation}
This implies that $G^{ij}$ and $G_{ij}$ for $i,j=1,2$ do not change the baryon number. 

The commutation relations among the fermionic and bosonic generators
are
\begin{equation}
\begin{split}
   [G^{ij},G^{k3}] =  \delta_{jk} G^{i3}, \qquad
   [G^{ij},G^{3k}] = - \delta_{ik} G^{3j}, \\
   [G_{ij},G_{k3}] =  \delta_{jk} G_{i3}, \qquad
   [G_{ij},G_{3k}] = - \delta_{ik} G_{3j}, \\
\end{split}
\end{equation}
and 
\begin{equation}
\begin{split}
   [G^{33},G^{i3}] = -  G^{i3}, \qquad
   [G^{33},G^{3i}] =  G^{3i}, \\
   [G_{33},G_{i3}] = -  G_{i3}, \qquad
   [G_{33},G_{3i}] =  G_{3i}.
\end{split}
\end{equation}
We can also show using the spin generators $G^{3}$ and $G_{3}$ that 
\begin{equation}
\begin{split}
  [G^{3}, G^{13}] = \frac{1}{2} G^{13}, \quad [G^{3}, G^{23}] = - \frac{1}{2} G^{23}, \quad
  [G^{3}, G^{31}] = -\frac{1}{2} G^{31}, \quad [G^{3}, G^{32}] =  \frac{1}{2} G^{32},\\
  [G_{3}, G_{13}] = \frac{1}{2} G_{13}, \quad [G_{3}, G_{23}] = - \frac{1}{2} G_{23}, \quad
  [G_{3}, G_{31}] = -\frac{1}{2} G_{31}, \quad [G_{3}, G_{32}] =  \frac{1}{2} G_{32}.
\end{split}
\end{equation}
These equations imply that $G^{13}$, $G^{32}$, $G_{13}$ and $G_{32}$ raise 
the 3rd component of spin by 1/2, while $G^{31}$, $G^{23}$, $G_{31}$ and $G_{23}$
lower by 1/2. The commutation relations with the baryon number operators $G^{8}$ and $G_{8}$
\begin{equation}
\begin{split}
  [G^{8}, G^{13}] = G^{13}, \quad [G^{8}, G^{23}] = G^{23}, \quad
  [G^{8}, G^{31}] = - G^{31}, \quad [G^{8}, G^{32}] =  - G^{23},\\
  [G_{8}, G_{13}] =  G_{13}, \quad [G_{8}, G_{23}] =   G_{23}, \quad
  [G_{8}, G_{31}] = - G_{31}, \quad [G_{8}, G_{32}] = - G_{23},
\end{split}
\end{equation}
show that $G^{13}$, $G^{23}$, $G_{13}$ and $G_{23}$ change 
the baryon number by 1, while $G^{31}$, $G^{32}$, $G_{31}$ and $G_{32}$
change the baryon number  by $-1$.

The anticommutation relations for the fermionic generators are
\begin{equation}
   \{G^{3i},G^{j3}\} = \delta_{ij} G^{33} + G^{ij}, \qquad
   \{G_{3i},G_{j3}\} = \delta_{ij} G_{33} + G_{ij},
\end{equation}
and the other commutation relations vanish:
\begin{equation}
   \{G^{i3},G^{j3}\} = \{G^{3i},G^{3j}\} = 0, \qquad
   \{G_{i3},G_{j3}\} = \{G_{3i},G_{3j}\} = 0.
\end{equation}
It is also notable that the relations
\begin{eqnarray}
   \{ G^{31}, G^{13}\} &=& G^{11} + G^{33} = \left[\frac{1}{2} (G^{11}+G^{22}) + G^{33} \right] + \frac{1}{2} (G^{11}-G^{22}) \\
   \{ G^{32}, G^{23}\} &=& G^{22} + G^{33} = \left[\frac{1}{2} (G^{11}+G^{22}) + G^{33} \right] - \frac{1}{2} (G^{11}-G^{22}) \\
   \ [ G^{21}, G^{12}] &=&  G^{11}-G^{22} 
\end{eqnarray}
implies that the commutation relations can only provide the combinations of the generators, 
$\sqrt\frac13(G^{11}+G^{22})+\sqrt\frac23 G^{33}$ and $\sqrt\frac12 (G^{11}-G^{22})$, 
but cannot $\sqrt\frac13(G^{11}+G^{22}-G^{33})$. This is true also for the generators with the subscript.

If we write the bosonic and fermionic generators as 
$B^{AB}$ and $F^{AB}$, respectively, that is,
\begin{equation}
  G^{11}, G^{22}, G^{12}, G^{21}, G^{33} \in B^{AB}, \qquad
  G^{13}, G^{23}, G^{31}, G^{32} \in F^{AB},
\end{equation}
the commutation relations are written as
\begin{align}
   [B^{AB}, B^{CD}] &= \delta_{BD} B^{AD} - \delta_{AD} B^{CB}, \\
   [F^{AB}, B^{CD}] & = \delta_{BC} F^{AD} - \delta_{AD} F^{CB}, \\
   \{ F^{AB}, F^{CD}\} &= \delta_{BC} B^{AD} + \delta_{AD} B^{CB}.
\end{align}
We have the same relations for the generators with the subscript.

As seen in the above commutation relations, $G^{AB}$ and $G_{AB}$ satisfy 
the same commutation relations. This implies that $G^{AB}$ and $G_{AB}$
are algebraically equivalent. According to Lorentz symmetry, both $\Psi^{i}$ and $\Psi_{i}$
fields should participate in the theory. 
Thus, in order to incorporate both $\Psi^{i}$ and $\Psi_{i}$ fields into the theory,
we would consider the V(3)$\otimes$V(3) symmetry. However, because the spin operator 
of the Dirac fermion in the standard notation is expressed as 
\begin{equation}
  {\bm S} =  \frac{1}{2} {\bm \Sigma} 
  = \frac{1}{2} \left( \begin{array}{cc} {\bm \sigma} & 0 \\ 0 & {\bm \sigma} \end{array}\right),
\end{equation}
which implies that the spin operation for the fermion acts on both $\psi^{i}$ and $\psi_{i}$
in the same direction, we should consider only the subalgebra of V(3)$\otimes$V(3)
which is generated by $\bm{G}_{AB} = G^{AB}+G_{AB}$.
As a result, $\bm{G}_{AB}$ satisfies the same commutation relations of $G^{AB}$ and $G_{AB}$, that is,
$\bm{G}_{AB}$ generates the V(3) algebra.

\section{Representations of V(3)}

\subsection{Fundamental representation: $\Psi^{i}\oplus \Psi_{i}$}
The triplets, $\Psi^{i}$ and $\Psi_{i}$, given in Eq.~\eqref{eq: pm field} are 
the fundamental representations of the V(3) algebra. 
These fields are transformed by the generators $G^{AB}$ and $G_{AB}$, respectively. 
Similarly the conjugate fields $\hat \Psi^{i}$ and $\hat \Psi_{i}$ are the complex representation 
of the fundamental representation.

The commutation relations of these fields for the spin $G^{3}$ and the baryon number $G^{8}$
read 
\begin{align}
   [G^{3}, \psi^{1}] &= - \frac{1}{2} \psi^{1}, & 
   [G^{3}, \psi^{2}] &= \frac{1}{2} \psi^{2}, & 
   [G^{3}, \varphi^{3}] &= 0, \\
   [G^{8}, \psi^{1}] &= - \frac{1}{3} \psi^{1}, & 
   [G^{8}, \psi^{2}] &= - \frac{1}{3} \psi^{2}, & 
   [G^{8}, \varphi^{3}] &= \frac{2}{3} \varphi^{3}.
\end{align}
This implies that $\psi^{1}$, $\psi^{2}$ and $\varphi^{3}$ fields have
the quantum number $(S_{3}, B)$ as $(1/2, 1/3)$, $(-1/2, 1/3)$ and $(0, -2/3)$, respectively. 

The same relations are satisfied for the fields with the subscript. 
Regarding $\Psi^{i} \oplus \Psi_{i}$ as a color triplet ``quark" field and
$\hat \Psi^{i} \oplus \hat \Psi_{i}$ as a ``antiquark" field with a color anti-triplet, 
we construct the representation of ``hadron'' as a composites of the quark fields.


With the commutation relations derived in the previous section,
we find that $\hat \Psi^{i} \Psi^{i}$ and $\hat \Psi_{i} \Psi_{i}$
are invariant under any transformations $G^{AB}$ and $G_{AB}$, 
respectively, as
\begin{equation}
   [G^{AB}, \hat \psi^{1}\psi^{1} + \hat \psi^{2} \psi^{2} 
   + \hat \varphi^{3} \varphi^{3} ] = 0, \qquad
   [G_{AB}, \hat \psi_{1}\psi_{1} + \hat \psi_{2} \psi_{2} 
   + \hat \varphi_{3} \varphi_{3} ] = 0.
\end{equation}
If we take a linear combination of these terms as $\hat \Psi^{i} \Psi^{i} - \hat \Psi_{i} \Psi_{i}$,
the mass term of Hamiltonian \eqref{eq:Hmass} is invariant 
under the V(3) transformation generated by $\bm{G}_{AB} = G^{AB}+G_{AB}$
as 
\begin{equation}
   [\bm{G}_{AB}, {\cal H}_{\rm mass}] 
   = [\bm{G}_{AB}, m (\hat \psi^{1}\psi^{1} + \hat \psi^{2} \psi^{2} 
   + \hat \varphi^{3} \varphi^{3})
   - m( \hat \psi_{1}\psi_{1} + \hat \psi_{2} \psi_{2} 
   + \hat \varphi_{3} \varphi_{3} )]
   = 0.
\end{equation}

\subsection{``Adjoint" Representation: $ \hat\Psi_i \Psi^j \oplus \hat \Psi^{i} \Psi_{j}$}
Next, we consider composite fields made of two fundamental representations, 
$\Psi^{i} \otimes \Psi_{i}$ and $\hat \Psi^{i} \otimes \hat \Psi_{i}$, which can be regarded
as ``mesonic" fields. We have two kinds of the combinations:
\begin{equation}
   \hat\Psi_i \Psi^j \oplus \hat \Psi^{i} \Psi_{j}, \qquad 
   \hat\Psi^{i} \Psi^j \oplus \hat \Psi_{i} \Psi_{j}.
\end{equation}
The former is favored by the ground state in the nonrelativistic limit, because $\Psi^{i}$ and $\hat \Psi_{i}$
contain the large components of the Dirac spinor. 
First, we consider the former combination. Let us introduce 
\begin{equation}
 	M_{i}^{\ j} = \hat\Psi_i \Psi^j ,\qquad
	M^{i}_{\ j} = \hat\Psi^i \Psi_j. \label{eq:meson}
\end{equation}
These fields are algebraically independent and belong to the same representations, as we shall see below.
For the Lorentz symmetry we need both fields in an appropriate combination. 

Let us see the irreducible representation for $M_{i}^{\ j}$ and $M^{i}_{\ j}$. 
We start with the $M_{1}^{\ 2}$ field, which has $S_{3}=-1$ for the 3rd component of spin 
and $B=0$ for the baryon number, and therefore $M_{1}^{\ 2}$ is a boson field. 
The generator ${\bm G}_{21}$ raises $1/2$ for the 3rd component of spin and does not change the baryon number. 
Having the commutation relations
\begin{equation}
	[\bm{G}_{21} , M_{1}^{\ 2}] = M_{2}^{\ 2} - M_{1}^{\ 1}, \qquad
	[\bm{G}_{21} , \frac1{\sqrt 2} (M_{1}^{\ 1} - M_{2}^{\ 2})] =  \sqrt 2 M_{2}^{\ 1}, \qquad
	[\bm{G}_{21} ,  M_{2}^{\ 1}] = 0,
\end{equation}
we find that $\{-M_{1}^{\ 2}, \frac{1}{\sqrt 2}(M_{1}^{\ 1} - M_{2}^{\ 2}), M_{2}^{\ 1}\}$ form a spin triplet. 
Thus, these fields have total spin 1 with $B=0$. We also find that  
the $\frac{1}{\sqrt 2}(M_{1}^{\ 1} + M_{2}^{\ 2})$ field orthogonal to $\frac{1}{\sqrt 2}(M_{1}^{\ 1} - M_{2}^{\ 2})$
has total spin 0. 

Next let us consider the transformation of $M_{1}^{\ 2}$ by ${\bm G}_{23}$ which 
raises the 3rd component of spin by $1/2$ and change the baryon number by $-1$:
\begin{equation}
   [{\bm G}_{23}, M_{1}^{\ 2}] = -M_{1}^{\ 3}.
\end{equation}
This implies that $M_{1}^{\ 3}$ has $S_3=-1/2$ and $B=-1$, and thus, it is a fermion field.
We consider the transformation of the field $M_{1}^{\ 3}$ by $\bm{G}_{21}$ changing $S_{3}$ by $+1$
as 
\begin{equation}
   [{\bm G}_{21}, M_{1}^{\ 3}] = M_{2}^{\ 3},
\end{equation}
which implies that $M_{2}^{\ 3}$ has $S_{3} = 1/2$ and $B=-1$. Thus, $\{ -M_{1}^{\ 3}, M_{2}^{\ 3}\}$
form a spin doublet with total spin $1/2$ and $B=-1$. Considering also the transformation 
of $M_{1}^{\ 2}$ by $\bm G_{31}$ which changes $S_{3}$ by $1/2$
and $B$ by $1$ as
\begin{equation}
  [\bm{G}_{31}, M_{1}^{\ 2}] = M_{3}^{\ 2}
\end{equation}
we find that the field $M_{3}^{\ 2}$ has $S_{3}=-1/2$ and $B=1$. The commutation relation 
\begin{equation}
  [ \bm{G}_{21}, M_{3}^{\ 2}] = -M_{3}^{\ 1}
\end{equation}
implies that $\{ -M_{3}^{\ 2}, M_{3}^{\ 1}\}$ form a spin doublet having total spin $1/2$ and baryon 
number $B=1$. These fields are fermions. 
Finally the commutation relation 
\begin{equation}
   \{\bm{G}_{23}, M_{3}^{\ 2}\} = M_{2}^{\ 2} + M_{3}^{\ 3} 
\end{equation}
shows that the field $M_{3}^{\ 3}$, which has total spin $S=0$ and baryon number $B=0$, 
is also within this multiplet. This implies that we need all of three components, 
$\frac{1}{\sqrt 2} ( M^{1}_{\ 1} - M^{2}_{\ 2})$, $\frac{1}{\sqrt 2} ( M^{1}_{\ 1} + M^{2}_{\ 2})$,
$M^{3}_{\ 3}$, to express $M_{2}^{\ 2} + M_{3}^{\ 3}$.

Similarly for the $M^{i}_{\ j}$ field, 
we find that $\{ - M^{2}_{\ 1}, \frac{1}{\sqrt 2} ( M^{1}_{\ 1} - M^{2}_{\ 2}), M^{1}_{\ 2}\}$
forms the spin triplet with $B=0$, $\frac 1{\sqrt 2}(M^{1}_{\ 1} + M^{1}_{\ 2})$ has total spin $S=0$
and baryon number $B=0$, $\{-M^{1}_{\ 3}, -M^{2}_{\ 3}\}$ and $\{-M^{3}_{\ 1}, - M^{3}_{\ 2}\}$ are
spin doublets with $B=-1$ and $B=1$, respectively, and $M^{3}_{\ 3}$ is the spin singlet with $B=0$.
We also confirm in the same way that $\hat\Psi^{i} \Psi^j \oplus \hat \Psi_{i} \Psi_{j}$ forms 
a nonet of V(3). 

In this way, we find a nonet representation of V(3) and write ${\bm 3} \otimes \bm {\bar{ 3}} = \bm 9$
instead of $\bm 3 \otimes \bm{\bar{ 3}} = \bm 1 \oplus \bm 8$ in SU(3).


\subsection{$\Psi^{i}\Psi^{j} \oplus \Psi_{i} \Psi_{j}$ representations}

We consider higher dimensional representations composed of two fundamental representations
$(\Psi^{i}\otimes \Psi^{j})\oplus(\Psi_{i}\otimes \Psi_{j})$, which can be regarded as ``diquark'' fields. 
Again, $\Psi^{i}\otimes \Psi^{j}$ is favored by the ground state in the nonrelativistic limit than 
$\Psi_{i}\otimes \Psi^{j}$. Here we consider the decomposition of $\Psi^{i}\otimes \Psi^{j}$
into the irreducible representations of V(3). The other combinations are also decomposed in the same way. 

We introduce nine fields, $\Psi^{i}_{a} \Psi^{j}_{b}$, where $i$ and $j$ are indices of the V(3)
fundamental representation running 1 to 3, and $a$ and $b$ are fixed labels representing other quantum 
numbers such as color.  The product $\Psi^{i} \Psi^{j}$ has 9 components. Here we consider 
the decomposition of the 9 components into the irreducible representations of V(3). 
We will see that $\Psi^{i} \Psi^{j}$ is decomposed into a quintet and a quartet representations,
that is written as $\bm 3 \otimes \bm 3 = \bm 5 \oplus \bm 4$.

\subsubsection{Quintet}
Let us start a highest field $\psi^1_a\psi^1_b$ which has  $S_3=+1$ and baryon number $B=2/3$. 
We lower spin quantum number by take the commutation relations
\begin{align}
	&[ G^{12} , \psi^1_a\psi^1_b] = -\psi^1_a\psi^2_b - \psi^2_a\psi^1_b \\
	&[ G^{12} , -(\psi^1_a\psi^2_b + \psi^2_a\psi^1_b)] = 2\psi^2_a\psi^2_b,
\end{align}
and then we find that 
$\{ \psi^2_a\psi^2_b , \frac{1}{\sqrt{2}}(\psi^1_a\psi^2_b + \psi^2_a\psi^1_b) , \psi^1_a\psi^1_b \}$ forms
a spin triplet with total spin~1 and baryon number $2/3$. 
Next we consider the transformation of the field $\psi^1_a\psi^1_b$ by $ G^{13}$,
\begin{align}
	[ G^{13} , \psi^1_a\psi^1_b] = \varphi^3_a\psi^1_b - \psi^1_a\varphi^3_b,
\end{align}
and consider the transformation of the field $\varphi^3_a\psi^1_b - \psi^1_a\varphi^3_b$ by $ G^{12}$ as 
\begin{align}
	[ G^{12} , \varphi^3_a\psi^1_b - \psi^1_a\varphi^3_b] = -\varphi^3_a\psi^2_b + \varphi^3_b\psi^2_a.
\end{align}
Therefore, we find that 
$\{\varphi^3_a\psi^2_b - \varphi^3_b\psi^2_a , \varphi^3_a\psi^1_b - \psi^1_a\varphi^3_b\}$ 
form a spin doublet with total spin $1/2$ and baryon number $-1/3$. 

Taking the transformation of the field $\varphi^3_a\psi^1_b - \psi^1_a\varphi^3_b$ by $ G^{13}$ as 
\begin{align}
	\{ G^{13} , \varphi^3_a\psi^1_b - \psi^1_a\varphi^3_b\} = 0,
\end{align}
we find that there are no components with baryon number $-4/3$ for the multiplet starting with $\psi^1_a\psi^1_b$.

Consequently, there are five components, 
\begin{equation}
\psi^2_a\psi^2_b ,\quad 
\frac{1}{\sqrt{2}}(\psi^1_a\psi^2_b + \psi^2_a\psi^1_b) , \quad
\psi^1_a\psi^1_b, \quad
\frac{1}{\sqrt{2}}(\varphi^3_a\psi^2_b - \varphi^3_b\psi^2_a), \quad 
\frac{1}{\sqrt{2}}(\varphi^3_a\psi^1_b - \psi^1_a\varphi^3_b)
\end{equation}
forming a quintet representation $\bm 5$. This representation is anti-symmetric under the exchange 
of indices $a$ and $b$. (Note that $\psi^{i}$ is a fermion field having $\{\psi^{i}_{a}, \psi^{j}_{b}\} = 0$.)

\subsubsection{Quartet}
To find further representation, 
we start with $\frac{1}{\sqrt{2}}(\psi^1_a\psi^2_b - \psi^2_a\psi^1_b)$ which is 
orthogonal to $\frac{1}{\sqrt{2}}(\psi^1_a\psi^2_b + \psi^2_a\psi^1_b)$ and has spin 0.
Calculating the commutation relations
\begin{align}
	[ G^{23} , \psi^1_a\psi^2_b - \psi^2_a\psi^1_b] &= -\psi^1_a\varphi^3_b -\varphi^3_a\psi^1_b, \\
	[ G^{12} , \psi^1_a\varphi^3_b +\varphi^3_a\psi^1_b] &=  \psi^2_a\varphi^3_b +\varphi^3_a\psi^2_b, \\
	\{ G^{13} , \psi^1_a\varphi^3_b +\varphi^3_a\psi^1_b\} &= 2\varphi^{3}_a\varphi^3_b , 
\end{align}
we find that four components,
\begin{equation}
  \frac{1}{\sqrt{2}}(\psi^1_a\psi^2_b - \psi^2_a\psi^1_b), \quad
  \frac{1}{\sqrt{2}}(\psi^2_a\varphi^3_b +\varphi^3_a\psi^2_b), \quad
  \frac{1}{\sqrt{2}}(\psi^1_a\varphi^3_b +\varphi^3_a\psi^1_b), \quad
  \varphi^{3}_a\varphi^3_b,  \label{eq:quartet}
\end{equation}
form a quartet representation $\bm 4$. The first and last terms in Eq.~\eqref{eq:quartet}
are spin singlets and have baryon number $2/3$ and $-4/3$, respectively, 
while the middle two terms form a spin doublet with baryon number $-1/3$.  
This representation is symmetric under the exchange of indices $a$ and $b$. 

As a result, we have the decomposition $\bm 3 \otimes \bm 3 = \bm 5_{\rm A} \oplus \bm 4_{\rm S}$,
where A and S stand for anti-symmetry and symmetry under the exchange of two fundamental 
representations.

\subsection{$\Psi^{i} \Psi^{j} \Psi^{k} \oplus \Psi_{i} \Psi_{j} \Psi_{k}$ representations}

We decompose the fields composed of three fundamental representations,
$\Psi^{i}_{a} \Psi^{j}_{b} \Psi^{k}_{c}$, into irreducible representations of V(3), 
where again $a$, $b$ and $c$ label other quantum numbers. This field configuration 
is favored by the ground state in the nonrelativistic limit and 
corresponds to baryons for color singlet. Finally we will find that $\Psi^{i}_{a} \Psi^{j}_{b} \Psi^{k}_{c}$
is decomposed into a septet, a quartet and two octet representations, namely we 
write $\bm 3 \otimes \bm 3 \otimes \bm 3 = \bm 7 \oplus \bm 4 \oplus \bm 8 \oplus \bm 8$. 
Other configurations, $\Psi_{i} \Psi^{j} \Psi^{k}$, $\Psi_{i} \Psi_{j} \Psi^{k}$ and $\Psi_{i} \Psi_{j} \Psi_{k}$
forms the same multiplets. 

\subsubsection{Septet}
We start with a highest component $\psi^1_a \psi^1_b \psi^1_c$ with spin $S_3=3/2$ and baryon number $B=1$. 
Taking the commutation relations
\begin{align}
	&[ G_{12} , \psi^1_a \psi^1_b \psi^1_c] = -\psi^2_a\psi^1_b\psi^1_c -\psi^1_a\psi^2_b\psi^1_c -\psi^1_a\psi^1_b\psi^2_c, \\
	&[ G_{12} , -\psi^2_a\psi^1_b\psi^1_c -\psi^1_a\psi^2_b\psi^1_c -\psi^1_a\psi^1_b\psi^2_c] = 2( \psi^2_a\psi^2_b\psi^1_c + \psi^2_a\psi^1_b\psi^2_c +\psi^1_a\psi^2_b\psi^2_c ), \\
	&[ G_{12} , \psi^2_a\psi^2_b\psi^1_c + \psi^2_a\psi^1_b\psi^2_c +\psi^1_a\psi^2_b\psi^2_c] = -3\psi^2_a\psi^2_b\psi^2_c,
\end{align}
we find a spin quartet 
$\{\psi^1_a \psi^1_b \psi^1_c, 
\frac 1 {\sqrt 3} (\psi^2_a\psi^1_b\psi^1_c +\psi^1_a\psi^2_b\psi^1_c +\psi^1_a\psi^1_b\psi^2_c),
\frac 1 {\sqrt 3}( \psi^2_a\psi^2_b\psi^1_c + \psi^2_a\psi^1_b\psi^2_c +\psi^1_a\psi^2_b\psi^2_c),
\psi^2_a\psi^2_b\psi^2_c\}$ with total spin $3/2$ and  baryon number $1$. 
Next, we consider the transformation of $\psi^1_a \psi^1_b \psi^1_c$ by $G_{13}$
and the spin partners of its product:
\begin{align}
	\{ G_{13} , \psi^1_a \psi^1_b \psi^1_c\} &=\varphi^3_a\psi^1_b\psi^1_c - \psi^1_a\varphi^3_b\psi^1_c + \psi^1_a\psi^1_b\varphi^3_c, \\
	[ G^{12} , \varphi^3_a\psi^1_b\psi^1_c - \psi^1_a\varphi^3_b\psi^1_c + \psi^1_a\psi^1_b\varphi^3_c] 
	&= -\varphi^3_a\psi^2_b\psi^1_c - \varphi^3_a\psi^1_b\psi^2_c  + \psi^2_a\varphi^3_b\psi^1_c +\psi^1_a\varphi^3_b\psi^2_c 
	\nonumber \\
	& \quad- \psi^2_a\psi^1_b\varphi^3_c - \psi^1_a\psi^2_b\varphi^3_c, \\
	[ G^{12} , -\varphi^3_a\psi^2_b\psi^1_c - \varphi^3_a\psi^1_b\psi^2_c + \psi^2_a\varphi^3_b\psi^1_c
	 \nonumber \\ 
	+\psi^1_a\varphi^3_b\psi^2_c - \psi^2_a\psi^1_b\varphi^3_c - \psi^1_a\psi^2_b\varphi^3_c]  
	&= 2( \psi^2_a\varphi^3_b\psi^2_c - \psi^2_a\psi^2_b\varphi^3_c - \varphi^3_a\psi^2_b\psi^2_c ).
\end{align}
These form spin triplet with baryon number $0$.
Decreasing the baryon number of these terms further,  we find
\begin{align}
	[ G^{13} , \varphi^3_a\psi^1_b\psi^1_c - \psi^1_a\varphi^3_b\psi^1_c + \psi^1_a\psi^1_b\varphi^3_c] = 0.
\end{align}
This implies that we have a septet representation $\bm 7$ as
\begin{gather}
\psi^1_a \psi^1_b \psi^1_c, \ 
\frac 1 {\sqrt 3} (\psi^2_a\psi^1_b\psi^1_c +\psi^1_a\psi^2_b\psi^1_c +\psi^1_a\psi^1_b\psi^2_c),\ 
\frac 1 {\sqrt 3}( \psi^2_a\psi^2_b\psi^1_c + \psi^2_a\psi^1_b\psi^2_c +\psi^1_a\psi^2_b\psi^2_c),\ 
\psi^2_a\psi^2_b\psi^2_c, \nonumber \\
\frac 1 {\sqrt 3} (\varphi^3_a\psi^1_b\psi^1_c - \psi^1_a \varphi^3_b\psi^1_c + \psi^1_a\psi^1_b\varphi^3_c),
\nonumber \\
\frac 1 {\sqrt 6} (\varphi^3_a\psi^2_b\psi^1_c + \varphi^3_a \psi^1_b\psi^2_c  - \psi^2_a\varphi^3_b\psi^1_c 
-\psi^1_a\varphi^3_b\psi^2_c + \psi^2_a\psi^1_b\varphi^3_c + \psi^1_a\psi^2_b\varphi^3_c), \nonumber \\
\frac 1 {\sqrt 3} ( \psi^2_a\varphi^3_b\psi^2_c - \psi^2_a \psi^2_b\varphi^3_c - \varphi^3_a\psi^2_b\psi^2_c )
\end{gather}
where the first four terms have total spin 3/2 and baryon number $1$ and 
the last three terms have total spin 1 and baryon number $0$. These terms are totally anti-symmetric 
under the exchange of indices $a$, $b$ and $c$. 

\subsubsection{Quartet}
To find further representations, we start with another highest component,
$\varphi^3_a\varphi^3_b\varphi^3_c$, which has spin 0 and baryon number $-2$. Increasing its baryon number by $ G^{31}$, we have 
\begin{align}
	[ G^{31} , \varphi^3_a\varphi^3_b\varphi^3_c] = -\psi^1_a\varphi^3_b\varphi^3_c - \varphi^3_a\psi^1_b\varphi^3_c - \varphi^3_a\varphi^3_b\psi^1_c,
\end{align}
which has spin $S_3=+1/2$ and baryon number $B=-1$, and its spin partner 
can be found by applying $G^{12}$ on it as
\begin{align}
	[ G^{12} , -\psi^1_a\varphi^3_b\varphi^3_c - \varphi^3_a\psi^1_b\varphi^3_c - \varphi^3_a\varphi^3_b\psi^1_c] = -\psi^2_a\varphi^3_b\varphi^3_c - \varphi^3_a\psi^2_b\varphi^3_c - \varphi^3_a\varphi^3_b\psi^2_c,
\end{align}
which has spin $S_3=-1/2$ and baryon number $B=-1$. Calculating  
\begin{align}
	\{ G^{32} ,-\psi^1_a\varphi^3_b\varphi^3_c - \varphi^3_a\psi^1_b\varphi^3_c - \varphi^3_a\varphi^3_b\psi^1_c\} 
	=\ &  \varphi^3_a\psi^2_b\psi^1_c + \psi^2_a\varphi^3_b\psi^1_c 
	-\psi^1_a\varphi^3_b\psi^2_c - \varphi^3_a\psi^1_b\psi^2_c
	\nonumber \\ &
	-\psi^1_a \psi^2_b \varphi^3_c +  \psi^2_a\psi^1_b\varphi^3_c,
\end{align}
we have a further component with spin 0 and baryon number 0 
in this multiplet. 
Thus the second multiplet of three fundamental representations is
a quartet representation $\bm 4$:
\begin{gather}
 \frac1{\sqrt{6}} (\varphi^3_a\psi^2_b\psi^1_c + \psi^2_a\varphi^3_b\psi^1_c 
	-\psi^1_a\varphi^3_b\psi^2_c - \varphi^3_a\psi^1_b\psi^2_c
	-\psi^1_a \psi^2_b \varphi^3_c +  \psi^2_a\psi^1_b\varphi^3_c )\\
 \frac 1{\sqrt 3}(\psi^1_a\varphi^3_b\varphi^3_c+\varphi^3_a\psi^1_b\varphi^3_c+ \varphi^3_a\varphi^3_b\psi^1_c), \quad
 \frac1{\sqrt 3}(\psi^2_a\varphi^3_b\varphi^3_c+\varphi^3_a\psi^2_b\varphi^3_c+ \varphi^3_a\varphi^3_b\psi^2_c), \\
 \varphi^3_a\varphi^3_b\varphi^3_c.
\end{gather}
This representation is symmetric 
under the exchange of indices $a$, $b$ and $c$. 

\subsubsection{Octet symmetric}
Next, we see another representation by starting with a spin double 
$\frac{1}{\sqrt{2}}(\psi^1_a\psi^2_b - \psi^2_a\psi^1_b)\psi^1_c$, 
$\frac{1}{\sqrt{2}}(\psi^1_a\psi^2_b - \psi^2_a\psi^1_b)\psi^2_c$
with baryon number 1, which are symmetric under the exchange of
indices $a$ and $b$.  We decrease the baryon number of the former term:
\begin{align}
	\left\{ G^{23} , \frac{1}{\sqrt{2}}(\psi^1_a\psi^2_b - \psi^2_a\psi^1_b)\psi^1_c\right\} = - \frac{1}{\sqrt{2}}(\psi^1_a\varphi^3_b\psi^1_c + \varphi^3_a\psi^1_b\psi^1_c).
\end{align}
This term has spin $S_3=+1$ and baryon number $B=0$. Its spin partners 
are found by decreasing its spin with $G^{12}$ sequently as
$\frac{1}{2}(\psi^2_a\varphi^3_b\psi^1_c + \psi^1_a\varphi^3_b\psi^2_{c} + \varphi^3_a\psi^2_b\psi^1_c + \varphi^3_a\psi^1_b\psi^2_c)$ and 
$\frac 1{\sqrt{2}}(\psi^2_a\varphi^3_b\psi^2_c + \varphi^3_a\psi^2_b\psi^2_c)$. Thus, these components form a spin triplet with baryon number $0$.
We further apply $G^{13}$, and we have
\begin{align}
	\left[ G^{13} , \frac{1}{\sqrt{2}}(\psi^1_a\varphi^3_b\psi^1_c + \varphi^3_a\psi^1_b\psi^1_c)\right] = -\frac{1}{\sqrt{2}}(2\varphi^3_a\varphi^3_b\psi^1_c -\psi^1_a\varphi^3_b\varphi^3_c - \varphi^3_a\psi^1_b\varphi^3_c) 
\end{align}
This has spin $S_3=+1/2$ and baryon number $B=-1$. Its spin partner 
is found as $\frac{1}{\sqrt{6}}(2\varphi^3_a\varphi^3_b\psi^2_c -\psi^2_a\varphi^3_b\varphi^3_c - \varphi^3_a\psi^2_b\varphi^3_c)$. 
In addition, we have 
\begin{align}
	\left\{  G^{13} ,  \frac{1}{\sqrt{2}}(\psi^1_a\psi^2_b - \psi^2_a\psi^1_b)\psi^1_c \right\} = \frac{1}{\sqrt{2}}(\psi^2_a\varphi^3_b\psi^1_c + 
	\varphi^3_a\psi^2_b\psi^1_c + 
	\psi^1_a\psi^2_b\varphi^3_c - \psi^2_a\psi^1_b\varphi^3_c),
\end{align}
which has spin 0 and baryon number $0$. This can be written 
as a linear combination of two components as
\begin{eqnarray}
  \lefteqn{
   \psi^2_a\varphi^3_b\psi^1_c +\varphi^3_a\psi^2_b\psi^1_c +  \psi^1_a\psi^2_b\varphi^3_c - \psi^2_a\psi^1_b\varphi^3_c} 
  && \nonumber \\
  &= &\frac12 [\psi^2_a\varphi^3_b\psi^1_c + \varphi^3_a\psi^2_b\psi^1_c 
  + \psi^1_a\varphi^3_b\psi^2_{c} + \varphi^3_a\psi^1_b\psi^2_c]
  \nonumber \\ &&
  + \frac12 [\psi^2_a\varphi^3_b\psi^1_{c} +\varphi^3_a\psi^2_b\psi^1_c
     - \psi^1_a\varphi^3_b\psi^2_c  - \varphi^3_a\psi^1_b\psi^2_c
     +2( \psi^1_a\psi^2_b\varphi^3_c -  \psi^2_a\psi^1_b\varphi^3_c)].
\end{eqnarray}
Thus $\frac1{\sqrt {12}} [\psi^2_a\varphi^3_b\psi^1_{c} +\varphi^3_a\psi^2_b\psi^1_c
     - \psi^1_a\varphi^3_b\psi^2_c  - \varphi^3_a\psi^1_b\psi^2_c
     +2( \psi^1_a\psi^2_b\varphi^3_c -  \psi^2_a\psi^1_b\varphi^3_c)]$
is also a component of the multiplet. Consequently we have 
eight components in this multiplet forming a octet $\bm 8$:
\begin{gather}
\frac{1}{\sqrt{2}}(\psi^1_a\psi^2_b - \psi^2_a\psi^1_b)\psi^1_c, \quad
\frac{1}{\sqrt{2}}(\psi^1_a\psi^2_b - \psi^2_a\psi^1_b)\psi^2_c, \\
\frac{1}{\sqrt{2}}(\psi^1_a\varphi^3_b\psi^1_c + \varphi^3_a\psi^1_b\psi^1_c), \qquad\qquad\qquad
\frac 1{\sqrt{2}}(\psi^2_a\varphi^3_b\psi^2_c + \varphi^3_a\psi^2_b\psi^2_c),
\\
\frac{1}{2}(\psi^2_a\varphi^3_b\psi^1_c + \psi^1_a\varphi^3_b\psi^2_{c} + \varphi^3_a\psi^2_b\psi^1_c + \varphi^3_a\psi^1_b\psi^2_c), 
\\
\frac1{\sqrt {12}} [\psi^2_a\varphi^3_b\psi^1_{c} +\varphi^3_a\psi^2_b\psi^1_c
     - \psi^1_a\varphi^3_b\psi^2_c  - \varphi^3_a\psi^1_b\psi^2_c
     +2( \psi^1_a\psi^2_b\varphi^3_c -  \psi^2_a\psi^1_b\varphi^3_c)]\\
     -\frac{1}{\sqrt{2}}(2\varphi^3_a\varphi^3_b\psi^1_c -\psi^1_a\varphi^3_b\varphi^3_c - \varphi^3_a\psi^1_b\varphi^3_c), \quad
     \frac{1}{\sqrt{6}}(2\varphi^3_a\varphi^3_b\psi^2_c -\psi^2_a\varphi^3_b\varphi^3_c - \varphi^3_a\psi^2_b\varphi^3_c)
\end{gather}

\subsubsection{Octet asymmetric}

Next we start with another spin $\frac{1}{2}$ doublet 
$
\frac{1}{\sqrt{6}} (2\psi^1_a\psi^1_b\psi^2_c - (\psi^1_a\psi^2_b + \psi^2_a\psi^1_b)\psi^1_c )$,  
$\frac{1}{\sqrt{6}}(2\psi^2_a\psi^2_b\psi^1_c - (\psi^1_a\psi^2_b + \psi^2_a\psi^1_b)\psi^2_c )
$ with baryon number $1$, which is asymmetric under the exchange of
indices $a$ and $b$.
We decrease the baryon number of the former component 
by considering the transformation of $G^{23}$:
\begin{align}
	\left\{G^{23} , \frac{1}{\sqrt{6}}\left(2\psi^1_a\psi^1_b\psi^2_c - (\psi^1_a\psi^2_b + \psi^2_a\psi^1_b)\psi^1_c \right) \right\} = \frac{1}{\sqrt{6}}(2\psi^1_a\psi^1_b\varphi^3_c + \psi^1_a\varphi^3_b\psi^1_c - \varphi^3\psi^1_b\psi^1_c),
\end{align}
which has spin $S_3=+1$ and baryon number $B=0$. 
Its spin partners are found by operating $G^{12}$ as 
$\frac{1}{\sqrt{12}}[
    \psi^2_a\varphi^3_b\psi^1_c -  \varphi^3_a\psi^2_b\psi^1_c
   + \psi^1_a\varphi^3_b\psi^2_c  - \varphi^3_a\psi^1_b\psi^2_c
  + 2(\psi^2_a\psi^1_b\varphi^3_c + \psi^1_a\psi^2_b\varphi^3_c)
]$, 
$ \frac{1}{\sqrt{6}}(2\psi^2_a\psi^2_b\varphi^3_c  
+ \psi^2_a\varphi^3_b\psi^2_c  - \varphi^3_a\psi^2_b\psi^2_c)$.
We further decrease the baryon number by using $G^{13}$:
\begin{align}
	\left[ G^{13} , \frac{1}{\sqrt{6}}(2\psi^1_a\psi^1_b\varphi^3_c + \psi^1_a\varphi^3_b\psi^1_c - \varphi^3\psi^1_b\psi^1_c) \right] =  \frac{1}{\sqrt{6}}(3\varphi^3_a\psi^1_b\varphi^3_c - 3\psi^1_a\varphi^3_b\varphi^3_c),
\end{align}
and its spin partner is found as $\frac{1}{\sqrt 2}(\varphi^3_a\psi^2_b\varphi^3_c - \psi^2_a\varphi^3_b\varphi^3_c)$.
These components have spin $1/2$ and baryon number $-1$.
We also calculate 
\begin{align}
   \lefteqn{
    \left\{G^{13} , \frac{1}{\sqrt{6}}\left(2\psi^1_a\psi^1_b\psi^2_c - (\psi^1_a\psi^2_b + \psi^2_a\psi^1_b)\psi^1_c \right) \right\} } & \nonumber\\
   & = \frac{1}{\sqrt{6}}(
    2\varphi^3_a\psi^1_b\psi^2_c - 2\psi^1_a \varphi^3_b \psi^2_c
    - \varphi^3_a\psi^2_b \psi^1_c - \psi^1_a\psi^2_b \varphi^3_c
    + \psi^2_a\varphi^3_b \psi^1_c - \psi^2_a\psi^1_b \varphi^3_c).
\end{align}
This component can be written as
\begin{align}
  \lefteqn{
      \psi^2_a\varphi^3_b \psi^1_c  - \varphi^3_a\psi^2_b \psi^1_c
     - 2\psi^1_a \varphi^3_b \psi^2_c + 2\varphi^3_a\psi^1_b\psi^2_c
    - \psi^1_a\psi^2_b \varphi^3_c- \psi^2_a\psi^1_b \varphi^3_c
    } &\nonumber \\
    & = -\frac12 [
     \psi^2_a\varphi^3_b\psi^1_c -  \varphi^3_a\psi^2_b\psi^1_c
   + \psi^1_a\varphi^3_b\psi^2_c  - \varphi^3_a\psi^1_b\psi^2_c
  + 2(\psi^2_a\psi^1_b\varphi^3_c + \psi^1_a\psi^2_b\varphi^3_c)
  ] \nonumber \\ 
 & \quad +\frac32 [ 
     \psi^2_a\varphi^3_b\psi^1_c -  \varphi^3_a\psi^2_b\psi^1_c
   - \psi^1_a\varphi^3_b\psi^2_c  + \varphi^3_a\psi^1_b\psi^2_c].
\end{align}
Thus, we have the following 8 components in this multiplet:
\begin{gather}
   \frac{1}{\sqrt{6}} (2\psi^1_a\psi^1_b\psi^2_c - (\psi^1_a\psi^2_b + \psi^2_a\psi^1_b)\psi^1_c ), \quad  
   \frac{1}{\sqrt{6}}(2\psi^2_a\psi^2_b\psi^1_c - (\psi^1_a\psi^2_b + \psi^2_a\psi^1_b)\psi^2_c ) 
   \\
   \frac{1}{\sqrt{12}}[
    \psi^2_a\varphi^3_b\psi^1_c -  \varphi^3_a\psi^2_b\psi^1_c
   + \psi^1_a\varphi^3_b\psi^2_c  - \varphi^3_a\psi^1_b\psi^2_c
  + 2(\psi^2_a\psi^1_b\varphi^3_c + \psi^1_a\psi^2_b\varphi^3_c)]
  \\
   \frac{1}{\sqrt{6}}(2\psi^1_a\psi^1_b\varphi^3_c + \psi^1_a\varphi^3_b\psi^1_c - \varphi^3_{a}\psi^1_b\psi^1_c), \qquad \qquad
 \frac{1}{\sqrt{6}}(2\psi^2_a\psi^2_b\varphi^3_c  
+ \psi^2_a\varphi^3_b\psi^2_c  - \varphi^3_a\psi^2_b\psi^2_c)
  \\
     \frac12(\psi^2_a\varphi^3_b\psi^1_c -  \varphi^3_a\psi^2_b\psi^1_c
   - \psi^1_a\varphi^3_b\psi^2_c  + \varphi^3_a\psi^1_b\psi^2_c)
   \\
    \frac{1}{\sqrt{2}}(3\varphi^3_a\psi^1_b\varphi^3_c - 3\psi^1_a\varphi^3_b\varphi^3_c),
   \quad
   \frac{1}{\sqrt 2}(\varphi^3_a\psi^2_b\varphi^3_c - \psi^2_a\varphi^3_b\varphi^3_c).
\end{gather}

Then we find that the baryonic representation is 
\begin{align}
	\bm 3 \otimes \bm 3 \otimes \bm 3 = \bm 7_{A} \oplus \bm 4_{S} \oplus \bm 8_{\rho} \oplus \bm 8_{\lambda}.
\end{align}
where subscrips $A$ and $S$ mean totally asymmetry and symmetry under the exchange of indices $a$, $b$ and $c$, respectively, while subscript $\rho$ and $\lambda$ stand for asymmetry and symmetry under the exchange of indices $a$ and $b$, respectively.

\section{Representations of hadrons}

\begin{table}[b]
\caption{Possible hadrons in the same multiplet of V(3). The lowest states in each category 
are considered. The wavefunction of the orbital motion is assumed to be symmetric, while
for the multiplets with asterisks, their symmetry property makes the orbital wavefunction asymmetric
with orbital or radial excitation. The number in the parenthesis denotes the considerable spin-parity $J^{p}$
of the state. For the excited state,  the possible spin $S$ not total spin $J$ is written. 
One understands
that $c$ is a charm quark as a representative of quarks in the outside of 
the V(3) multiplets. One can replace the charm quark into another quark such as a bottom quark. 
}  \label{tab:multiplets}
\begin{center}
\begin{tabular}{cccc}
\hline\hline
  &  V(3)  & quark contents & hadrons \\
\hline
 $\hat \Psi c$ & $ \bar{\bm 3} $ 
 & $\bar s c$ ($1^{-}$, $0^{-}$), \quad $udc$ ($\frac 12^{+}$) 
   & $D_{s}^{*}$, $D_{s}$, $\Lambda_{c}$ \\
 $\hat \Psi \Psi$ & $\bm 9$ & $\bar s s$ ($1^{-}$, $0^{-}$), \quad $uds$ ($\frac12^{+}$), 
   \quad $\overline{ud}\bar s$ ($\frac12^{+}$),\quad $\overline{ud} ud$ ($0^{+}$) 
   & $\phi$, $\eta_{s}$, $\Lambda$, $f_{0}$ \\
 $\Psi \Psi c$ & $\bm 5$ &  $ssc$ ($\frac 12^{+}$, $\frac32^{+}$),\quad $\overline{ud}sc$ ($0^{+}$, $1^{+}$)
   &$\Omega_{c}$, $T_{cs}$ \\
  & $\bm 4^{*}$ & $ssc$ ($S=\frac12$),\quad $\overline{ud}sc$  ($S=0,\, 1$),\quad
  $\overline{ud}\overline{ud} c$ ($S=\frac12)$ & $\Omega_{c}$, $T_{cs}$, $\Theta_{c}$ \\
  $\Psi\Psi\Psi$ & $\bm 7$ & $sss$ ($\frac32^{+}$), \quad $\overline{ud}ss$ ($1^{+}$) & $\Omega$, $T_{ss}$ \\
  & $\bm 8^{*}$ & $sss$ ($S=\frac12$), \quad $\overline{ud}ss$ ($S=0,\, 1$), 
  \quad $\overline{ud}\overline{ud}s$ ($S=\frac12$) & $\Omega^{*}$, $T_{ss}$, $\Theta$ \\
  & $\bm 4^{**}$ & $\overline{ud}ss$ ($S=0$), \quad $\overline{ud}\overline{ud} s$ ($S=\frac12$), 
  \quad $\overline{ud}\overline{ud} \overline{ud} $ ($S=0$)& $T_{ss}$, $\Theta^{*}$, dibaryon \\
\hline\hline
\end{tabular}
\end{center}
\end{table}

Here we show examples of the V(3) representations for hadrons. 
Regarding that the strange constituent quark and the $ud$-diquark have a very similar mass,
such as 500 MeV, we assign the fundamental representation of V(3), $\bm 3$, into 
a strange quark with spin up, $s_{\uparrow}$, a strange quark with spin down, $s_{\downarrow}$, and 
an $ud$ antidiquark, $\overline{ud}$, as
 \begin{align}
 	\Psi^{i} = 
 	\begin{pmatrix}
 		\psi^{1} \\
 		\psi^{2} \\
 		\varphi^{3}
 	\end{pmatrix}
	=
 	\begin{pmatrix}
 		s_\uparrow \\
 		s_\downarrow \\
 		\overline{ud}
 	\end{pmatrix}, \label{eq:triplet}
 \end{align}
which has color triplet. We compose hadrons out of the $\Psi$ fields. 
The possible hadrons are summarized in Table~\ref{tab:multiplets}. 
We consider nonrelativisic favor configurations made of $\Psi^{i}$ and $\hat \Psi_{i}$.
It is easy to recover the relativistic covariance by making up other components 
composed of $\Psi_{i}$ and $\hat \Psi^{i}$ appropriately.
 
\subsection{Triplet Representation} \label{sec: triplet rep}
Since the triplet field $\Psi$ has color, a single $\Psi$ cannot form hadrons. 
We consider color singlet hadrons made of an heavy-quark and the anti-triplet field.
First we take the charm quark as one of the heavy quarks and consider hadrons composed 
of $c^{i}$ and $\hat \Psi_{j}$. The charm quark has spin 1/2 and $i$ runs from 1 to 2. 
Here we have six components, $\hat \psi_{1} c^{1}$, $\hat \psi_{2} c^{1}$, $\hat \varphi_{3} c^{1}$,
$\hat \psi_{1} c^{2}$, $\hat \psi_{2} c^{2}$, $\hat \varphi_{3} c^{2}$, which are two triplets of V(3).
For the spin eigenstates, we have a spin triplet, a spin singlet and a spin doublet as
\begin{equation}
 D^{*}_{s} = \{ \hat \psi_{2} c^{1}, \frac 1{\sqrt{2}}(\hat \psi_{1} c^{1} - \hat \psi_{2} c^{2}), \hat\psi_{1} c^{2} \},\  
 D_{s} = \frac 1{\sqrt{2}}(\hat \psi_{1} c^{1} + \hat \psi_{2} c^{2}),\ 
 \Lambda_{c} = \{\hat \varphi_{3} c^{1}, \hat \varphi_{3} c^{2}\}.
\end{equation}
Here we assign possible hadrons which have the appropriate quantum number. 
In this way, in the V(3) symmetry, $D^{*}_{s}$, $D_{s}$ and $\Lambda_{c}$ are in the same multiplet. 

We Introduce the conjugate fields like Eqs.~(\ref{eq:conjugate fermion}) and (\ref{eq:conjugate boson}) as,
\begin{equation}
 \hat D^{*}_{s} = \{ \hat  c^{1} \psi_{2}, \frac 1{\sqrt{2}}(\hat  c^{1} \psi_{1}- \hat  c^{2}\psi_{2}), 
  \hat c^{2} \psi_{1} \},\  
 \hat D_{s} =  \frac 1{\sqrt{2}}(\hat c^{1} \psi_{1} + \hat c^{2} \psi_{2}),\ 
 \hat \Lambda_{c} = \{-\hat c^{1} \varphi_{3}, -\hat c^{2} \varphi_{3}\}.
\end{equation}
Here we have used 
$(\hat \psi_{2} c^{1})^{\dagger} = (c^{1})^{\dagger}(\hat \psi_{2})^{\dagger} = \hat c^{1} \psi_{2}$ and
$(\hat \varphi_{3} c^{1})^{\dagger} = (c^{1})^{\dagger} (\hat \varphi_{3})^{\dagger} 
= -\hat c^{1} \varphi_{3}$ with $\hat \varphi_{3} = - \varphi_{3}^{\dagger}$.
The mass term for the V(3) triplet hadron can be written with a common mass $m_{0}$ as
\begin{align}
   \lefteqn{m_{0}(\hat D_{s}^{*} D_{s}^{*} + \hat D_{s} D_{s} + \hat \Lambda_{c} \Lambda_{c})} & \nonumber \\
   &= m_{0} \left[\hat c^{1} (\psi_{1} \hat \psi_{1} + \psi_{2} \hat \psi_{2} - \varphi_{3} \hat \varphi_{3})c^{1}
   +\hat c^{2} (\psi_{1} \hat \psi_{1} + \psi_{2} \hat \psi_{2} - \varphi_{3} \hat \varphi_{3})c^{2} \right]\\
   &= -m_{0} \left[\hat c^{1} (\hat \psi_{1} \psi_{1} + \hat \psi_{2} \psi_{2} + \hat \varphi_{3} \varphi_{3})c^{1}
   +\hat c^{2} (\hat \psi_{1} \psi_{1} + \hat \psi_{2} \psi_{2} + \hat \varphi_{3} \varphi_{3})c^{2} \right].
\end{align}
This is invariant under the V(3) rotation. Thus, if we have the V(3) symmetry, the masses of $D^{*}_{s}$,
$D_{s}$ and $\Lambda_{c}$ get degenerate. 

In the same way, if we consider a bottom quark $b$, 
$B^{*}_{s}$, $B_{s}$ and $\Lambda_{b}$ are in the same multiplet of V(3) as
\begin{equation}
 B^{*}_{s} = \{ \hat \psi_{2} b^{1}, \frac 1{\sqrt{2}}(\hat \psi_{1} b^{1} - \hat \psi_{2} b^{2}), \hat\psi_{1} b^{2} \},\  
 B_{s} = \frac 1{\sqrt{2}}(\hat \psi_{1} b^{1} + \hat \psi_{2} b^{2}),\ 
 \Lambda_{b} = \{\hat \varphi_{3} b^{1}, \hat \varphi_{3} b^{2}\}.
\end{equation}


\subsection{Mesonic nonet representation}
\label{sec:nonet}
Next let us consider the mesonic nonet representation, $\hat \Psi_{i} \Psi^{j} \oplus \hat \Psi^{i}\Psi_{j}$.
We introduce the matrix representation like Eq.~\eqref{eq:meson} as
\begin{equation}
   M_{i}^{\ j} = \hat \Psi_{i} \Psi^{j}.
\end{equation}
Considering the quantum number of $M_{i}^{\ j}$, we assign the following hadrons to the field $M_{i}^{\ j}$:
\begin{gather}
  \phi = \{ M_{2}^{\ 1}, \frac 1 {\sqrt 2} (M_{1}^{\ 1} - M_{2}^{\ 2}), M_{1}^{\ 2}\},\qquad
  \eta_{s} =  \frac{1}{\sqrt{2}}(M_{1}^{\ 1} +M_{2}^{\ 2}), \qquad
  f_{0} = M_{3}^{\ 3},  \nonumber \\
  \Lambda = \{M_{3}^{\ 1}, M_{3}^{\ 2}\}, \qquad
  \bar \Lambda = \{M_{2}^{\ 3}, M_{1}^{\ 3}\}. 
\end{gather}
Note that the $\hat \Psi_{i} \Psi_{j}$ combination has a pseudoscalar meson and a vector meson. 
Thus, $\phi$, $\eta_{s}$, $f_{0}$ and $\Lambda$ are in the same multiplet of V(3).
Here $\eta_{s}$ denotes the pseudoscalar meson composed of the strange quarks.

Introducing the conjugate fields given in Eqs.~(\ref{eq:conjugate fermion}) and (\ref{eq:conjugate boson}),
we have 
\begin{gather}
  \hat \phi = \{ M^{1}_{\ 2}, \frac 1 {\sqrt 2} (M^{1}_{\ 1} - M^{2}_{\ 2}), M^{2}_{\ 1}\},\qquad
  \hat \eta_{s} =  \frac{1}{\sqrt{2}}(M^{1}_{\ 1} +M^{2}_{\ 2}), \qquad
  \hat f_{0} = -M^{3}_{\ 3},  \nonumber \\
  \hat \Lambda = \{-M^{1}_{\ 3}, -M^{2}_{\ 3}\}, \qquad
  \hat {\bar \Lambda} = \{M^{3}_{\ 2}, M^{3}_{\ 1}\}. 
\end{gather}
Here we consider the $(+)$ component for these hadrons and we have used 
\begin{equation}
   \hat M_{i}^{\ j} = (M_{i}^{\ j})^{\dagger} = (\hat \Psi_{i} \Psi^{j})^{\dagger}
   = (\Psi^{j})^{\dagger} (\hat \Psi_{i})^{\dagger} = (-)^{\delta_{3i}} \hat \Psi^{j} \Psi_{i} 
   \equiv (-)^{\delta_{3i}}  M^{j}_{\ i}.
\end{equation}
The mass term can be read with a common mass $m_{0}$ as
\begin{align}
  \lefteqn{  m ( \hat \phi \phi + \hat \eta_{s} \eta_{s} + \hat f_{0} f_{0} + \hat \Lambda \Lambda + \hat {\bar \Lambda} \bar \Lambda) } & \nonumber \\
  &= m_{0} ( M^{1}_{\ 2} M_{2}^{\ 1} + M^{1}_{\ 1} M_{1}^{\ 1} + M^{2}_{\ 2} M_{2}^{\ 2} + M^{2}_{\ 1} M_{1}^{\ 2}
  - M^{3}_{\ 3} M_{3}^{\ 3} \nonumber \\ & \qquad
  - M^{1}_{\ 3} M_{3}^{\ 1} - M^{2}_{\ 3} M_{3}^{\ 2} + M^{3}_{\ 2} M_{2}^{\ 3} + M^{3}_{\ 1} M_{1}^{\ 3}) \\
  &= m_{0} \left[ \hat \psi^{1} (\psi_{1} \hat \psi_{1} + \psi_{2} \hat \psi_{2} - \varphi_{3} \hat \varphi_{3}) \psi^{1}
  + \hat \psi^{2} (\psi_{1} \hat \psi_{1} + \psi_{2} \hat \psi_{2} - \varphi_{3} \hat \varphi_{3}) \psi^{2}
  \right. \nonumber \\ & \qquad \left. 
  + \hat \varphi^{3} (\psi_{1} \hat \psi_{1} + \psi_{2} \hat \psi_{2} - \varphi_{3} \hat \varphi_{3}) \varphi^{3}
  \right]
\end{align}
This is again invariant under the V(3) transformation. Therefore the masses of
$\phi$, $\eta_{s}$, $f_{0}$ and $\Lambda$ get degenerate if we have the V(3) symmetry. 

For later convenience, we introduce a matrix form of these hadrons as
\begin{align}
	M &= 
	\begin{pmatrix}
		\frac{1}{\sqrt{2}}(\phi_0 + \eta_{s}) & \phi_\downarrow & \overline \Lambda_\downarrow \\
		\phi_\uparrow & -\frac{1}{\sqrt{2}}(\phi_0 - \eta_{s}) & \overline \Lambda_\uparrow \\
		\Lambda_\uparrow & \Lambda_\downarrow & f_0
	\end{pmatrix}, \quad
	\hat M = 
	\begin{pmatrix}
		\frac{1}{\sqrt{2}}(\hat \phi_0 + \hat \eta_{s}) & \hat \phi_\uparrow & -\hat \Lambda_\uparrow \\
		\hat \phi_\downarrow & -\frac{1}{\sqrt{2}}(\hat \phi_0 - \hat \eta_{s}) & -\hat \Lambda_\downarrow \\
		\hat{\overline{\Lambda}}_\downarrow & \hat{\overline{\Lambda}}_\uparrow & -\hat f_0
	\end{pmatrix}
\end{align}
The mass term can be written in the matrix form as
\begin{equation}
   m_{0} {\rm Tr} [\hat M M]. \label{eq:nonet mass}
\end{equation}

For the configuration $\hat \Psi^{i} \Psi^{j} \oplus \hat \Psi_{i} \Psi_{j}$, the multiplet is same 
but the parity is opposite. In this multiplet there are scalar and axial vector mesons instead of 
pseudoscalar and vector mesons for the configuration $\hat \Psi_{i} \Psi^{j} \oplus \hat \Psi^{i} \Psi_{j}$.
In the terminology of the nonrelativistic quark model, these hadrons are $p$-wave hadrons due to 
the orbital excitation of $\Psi$.

\subsection{Other representations}

Here we consider the other representations which we do not discuss their mass terms. 

\subsubsection{$\Psi\Psi c$ hadron}

Here we consider hadrons made of $\Psi \Psi c$. For the lowest states, the orbital wavefunction is symmetric. 
Since $c$ and $\Psi$ are color triplet, for the color single hadron states, the color configuration 
for the ``diquark" $\Psi\Psi$ should be anti-symmetric. Thus, $\Psi\Psi$ forms a quintet of V(3). 
In the quintet, we have $ss$ with spin 1 and $\overline{ud} s$ with spin 1/2. With the charm quark, 
in this representation we have
charmed baryons, $ssc$, with spin $J=1/2$ and $3/2$, which are $\Omega_{c}$,
and charmed tetraquarks, $\overline{ud} sc$, with spin $J=0$ and $1$.   
The $\overline{ud} sc$ meson is a genuin tetraquark, because all four quarks have a different flavor. Similarly, with the bottom quark, the $\Omega_{b}$ barions with spin 1/2 and 3/2 and tetraquark $\overline{ud} sb$ with spin 0 and 1 form a quintet. Later we will discuss the masses of these tetraquarks. The existence and stability of the tetraquark $\bar u \bar d cb$ was discussed in Ref.~\cite{Lee:2009rt}.

Allowing excitation of one $\Psi$, one can take the quartet representation of V(3) for the $\Psi\Psi$ configuration. 
With a $c$ quark, one has $css$ with spin $S=1/2$, $\overline{ud}sc$ with spin $S=0$ and $1$, 
and a pentaquark $\overline{ud} \overline{ud} c$ with spin $S=1/2$. Here we mention only the spin $S$ 
for these hadron, which can be fixed by the V(3) symmetry.
Orbital excitation brings an angular momentum into the system and 
we cannot fix the total spin $J$ before specifying the angular momentum. 
To determine the angular momentum, we should fix the details of the orbital wavefunction, 
which is beyond simple symmetry argument.

\subsubsection{Baryonic representation}

For the baryonic representation, we consider the $\Psi \Psi \Psi$ configuration. For the color singlet hadrons,
the three fields should be totally antisymmetric. For the lowest state in which the orbital wavefunction
is symmetric, the septet representation~$\bm 7$ of V(3) can be assigned.  In this representation, 
there are the $\Omega$ baryon composed of $sss$ with spin $J=3/2$ and a tetraquark made of $\overline{ud}ss$
with spin $J=1$. 

If one allows asymmetry in the orbital wavefunction with excitation of one of the $\Psi$ fields, 
the octet representation $\bm 8$ of V(3) is also possible. In this multiplet, we have
a exited $\Omega$ baryon, $sss$, with spin $S=1/2$, tetraquarks, $\overline{ud}ss$, with spin $S=0$ and $1$
and a pentaquark~$\Theta$, $\overline{ud}\overline{ud}s$,  with spin $S=1/2$. 
Again here we mention spin $S$ for these hadron,
because the angular momentum is not fixed. 

If one accepts two excitations, one can take the quartet representation $\bm 4$ of V(3), 
which is totally symmetric in the exchange of the $\Psi$ field. In this representation, there are 
an exited tetraquark $\overline{ud}ss$ with spin $S=0$, an excited pentaquark $\overline{ud}\overline{ud}s$
with spin $S=1/2$ and a dibaryon $\overline{ud}\overline{ud}\overline{ud}$ with spin $S=0$.






\section{Symmetry Breaking}
So far we have discussed the classifications of hadrons into the V(3) multiplets. 
The symmetry between the $s$ quark and the $\overline{ud}$ diquark is not fundamental
and should be broken by their mass difference and 
{spin-dependent}
interactions. 
{The spin dependent interactions also break the symmetry, because the $s$ quark and the $\overline{ud}$ diquark have a different spin.}
Here we consider 
the symmetry breaking effects on the degenerate mass of the hadrons in the same multiplet. 

For the fundamental representation~\eqref{eq:triplet}, since one has the spin symmetry between 
the first and second components, $\psi^{1}$ and $\psi^{2}$, the degeneracy of these components 
is exact.  Because there is no constraint by spin symmetry on the third component $\varphi^{3}$,
the mass degeneracy for $\varphi^{3}$ can be broken. Thus, we may write the mass term for
the $\Psi^{i}$ field as
\begin{equation}
   m_{0} \hat \Psi \Psi + \delta m \hat \Psi \lambda_{8} \Psi
\end{equation}
with a common mass $m_{0}$ and a parameter $\delta m$ representing the mass difference. 
Here $\lambda_{8}$ is the eighth component of the Gell-Mann matrix in the space of 
the triplet $\Psi$. In this way, the symmetry breaking of the mass of the fundamental representation 
can be expressed as $\lambda_{8}$ in the same way as the symmetry breaking on the quark masses
for the flavor SU(3).


\subsection{Triplet representation}
As seen in Sec.~\ref{sec: triplet rep}, $\{D_{s}^{*},\, D_{s},\, \Lambda_{c}\}$ form a triplet $\bar {\bm 3}$ of V(3)
and get degenerate in the symmetric limit. The symmetry breaking 
{stems from the mass difference between the $\bar s$~quark and the $ud$~diquark
and the spin-spin interaction between the $\bar s$ and $c$ quarks. The breaking term induced by 
the mass difference}
can be introduced by $\lambda_{8}$
in the V(3) space. Thus, we may write the mass term for this multiplet
$\Pi = \{D_{s}^{*},\, D_{s},\, \Lambda_{c}\}$ as
\begin{equation}
   m_{0} \hat \Pi\, \Pi + \delta m \hat \Pi \lambda_{8} \Pi
\end{equation}
with a common mass $m_{0}$ and a strength of the symmetry breaking $\delta m$ that introduces the mass
difference in the multiplet. From this mass term, we find the  masses of the hadrons in the multiplet as
\begin{equation}
    m_{D_{s}} = m_{D_{s}^{*}} = m_{0} + \frac{\delta m}{\sqrt 3}, \qquad
    m_{\Lambda_{c}} = m_{0} - \frac{2\delta m}{\sqrt 3}.
\end{equation}
{
The degeneracy of $D_{s}$ and $D_{s}^{*}$ is resolved by introduction of 
the spin-spin interaction between the $\bar s$ and $c$ quarks. 
}
With the spin-spin interaction, we have three parameters
which are not fixed by the symmetry argument for the three hadrons. Therefore, the V(3) symmetry breaking 
gives us no constraint among these masses. 
{Nevertheless, it is very interesting to point out that the mass difference between the baryon and the mesons stems from the mass difference of the $\bar s$ quark and the $ud$ diquark in our study, and, thus, this mass difference should be insensitive to the heavy quark physics. On the other hand, the $D_{s}$-$D_{s}^{*}$ mass splitting comes from the spin-spin interaction induced by the color magnetic force. The color magnetic interaction is proportional to the quark mass inverse. Thus, for the heavier quark the mass difference is more suppressed. This is known as the heavy quark spin symmetry and its breaking. Certainly if one compares the observed mass difference patterns in $(D_{s}^{*}, D_{s},\Lambda_{c})$ and $(B_{s}^{*}, B_{s},\Lambda_{b})$, one finds that the mass differences between the baryon and the spin averaged meson is independent of the heavy quark flavor. Note that one has to take the spin averaged mass for the mesons to remove the effect of the spin-spin interaction. In addition, as well know as the heavy quark symmetry, the mass differences of the mesons are suppressed for the heavier quark flavor. This is consistent with what we have seen in the present study.}


\subsection{Mesonic nonet representation}

\subsubsection{Mass formula}
In Sec.~\ref{sec:nonet}, we have discussed that $\{ \phi, \eta_{s}, f_{0}, \Lambda\}$ form a nonet 
representation of V(3) and get degenerate in the symmetric limit. 
The symmetric mass term of this multiplet is written as Eq.~\eqref{eq:nonet mass} in the matrix form. 
Now let us introduce the V(3) breaking effect on the mass term as
\begin{equation}
  m_{0} {\rm Tr}[\hat M M ] + \delta m {\rm Tr} [\hat M \lambda_{8} M]
\end{equation}
with a common mass $m_{0}$ and a mass difference parameter $\delta m$. Thanks to 
the properties $\hat M = M^{\dagger}$ and $\lambda_{8}^{\dagger} = \lambda_{8}$, 
we have ${\rm Tr} [\hat M \lambda_{8} M] = {\rm Tr}[M \lambda_{8} \hat M]$. Thus,
the V(3) breaking term by $\lambda_{8}$ is represented by a single parameter. 
Here we have not introduced the spin-spin interaction between the strange quarks, which 
is another source of the V(3) symmetry breaking. 

Calculating the mass term, we obtain
\begin{equation}
\begin{split}
 \lefteqn{
  m_{0} {\rm Tr}[\hat M M ] + \delta m {\rm Tr} [\hat M \lambda_{8} M]
  } & \\
  &= \left(m_0 + \frac{\delta M}{\sqrt{3}}\right)(\hat \phi \phi +\hat \eta_{s} \eta_{s})
  + \left(m_0 - 2\frac{\delta M}{\sqrt{3}}\right)\hat f_0 f_0 
  + \left(m_0 - \frac{\delta M}{2\sqrt{3}}\right)(\hat \Lambda \Lambda +\hat{\overline{\Lambda}} \overline{\Lambda})
\end{split}
\end{equation}
This implies that the masses of these hadrons are obtained as
\begin{equation}
   m_\phi = m_{\eta_{s}} = m_0 + \frac{\delta m}{\sqrt 3}, \qquad
   m_{f_0} = m_0 - \frac{2\delta m}{\sqrt 3}, \qquad
   m_\Lambda = m_0 - \frac{\delta m}{2 \sqrt 3}.
\end{equation}
The degeneracy between $\phi$ and $\eta_{s}$ can be resolved by introducing the spin-spin 
interaction between the $s$ quarks. Eliminating the parameters $m_{0}$ and $\delta m$ in 
these mass formulae, we obtain a mass relation 
\begin{equation}
	2m_{\Lambda} = m_{f_0} + m_{\bar ss},
	\label{eq:mass-rel}
\end{equation}
where we have introduced $m_{\bar ss} = m_{\phi} = m_{\eta_{s}}$. 
This is a Gell-Mass Okubo mass formula for the V(3) symmetry. 

\subsubsection{Discussion}
Here let us discuss how the symmetry property of V(3) works in the mass formula~\eqref{eq:mass-rel} 
for the $\phi, \eta_{s}, f_{0}, \Lambda$ hadrons. 
First of all, in the mass formula, we do not take account of the spin-spin interaction between the 
strange quarks which induces the mass spitting between the spin partners, $\phi$ and $\eta_{s}$.
A simple way to resolve the spin-spin splitting is to take a spin average 
\begin{equation}
   m_{\bar ss} = \frac 14 (3 m_{\phi} + m_{\eta_{s}}).
\end{equation}
This is obtained accordingly to the first order perturbative calculation, in which
the mass shifts induced by the spin-spin interaction are given as
\begin{equation}
   m_{\phi} = m_{\bar ss} + \frac14 \alpha_{\bar ss}, \qquad m_{\eta_{s}} = m_{\bar ss} - \frac 34 \alpha_{\bar ss}, 
\end{equation}
for the spin 1 and 0 states, respectively, with a strength parameter $\alpha_{\bar ss}$ of the spin splitting for the strange quarks. 
Nevertheless, we do not use the physical $\eta$ mass to resolve the spin splitting due to the following reason;
For the vector meson, it is well-known that the flavor SU(3) breaking induces the mixing between the flavor octet
and singlet with isospin $I=0$ and strangeness $S=0$ and the quark contents of the $\omega$ and $\phi$ mesons
are written with the ideal mixing in which the $\omega$ meson and
the $\phi$ meson may be composed of $\frac 1 {\sqrt 2} (\bar uu + \bar dd)$ and $\bar ss$, respectively. 
Therefore, the quark content of the $\phi$ meson in our picture is consistent with the physical $\phi$ meson. 
For the pseudoscalar meson, however, the mixing between the flavor octet and singlet is 
known to be substantially small due to the different origin of the masses for the octet and 
singlet pseudoscalar mesons. Hence, the flavor content of the physical $\eta$ meson is 
almost given by the octet representation of the flavor SU(3), that is
$\frac 1{\sqrt 6}(\bar uu + \bar dd - 2 \bar ss)$, not purely $\bar ss$.
This is not consistent with our picture and we cannot directly apply the physical $\eta$ mass to 
our mass formula. 

Here we estimate the magnitude of the spin-spin splitting of $\bar ss$
in the following way. The spin-spin interaction is induced by the color magnetic interaction 
of the one gluon exchange. The strength of the color magnetic interaction is in inverse proportion to 
the masses of the quarks participating in the spin-spin interaction.
Thus, the strength parameter $\alpha_{\bar ss}$ for the strange quarks may be written as
\begin{equation}
   \alpha_{\bar ss} = \frac{\beta}{m_{s}^{2}},
\end{equation}
where $\beta$ is a universal parameter of the spin-spin interaction independent of 
the flavor of the participating quarks. 
Here we use the spin-spin splitting between $D^{*}_{s}$ and $D_{s}$
composed of the $c$ and $\bar s$ quarks, and 
the observed mass splitting is found to be $0.1438 \pm 0.0004$ GeV~\cite{Tanabashi:2018oca}.
The mass different can be written as
\begin{equation}
   m_{D^{*}_{s}} - m_{D_{s}} = \alpha_{\bar s c} = \frac{\beta}{m_{s} m_{c}}.
\end{equation}
Taking the observed mass difference as 0.14 GeV and assuming that 
the masses of the strange and charm quarks be 0.5 GeV and 1.3 GeV, respectively, 
we obtain $ \beta = 0.094$ GeV$^{3}$. With this value we also 
reproduce the mass different of $B_{s}^{*}$ and $B_{s}$ as
0.047 GeV for the $b$ quark mass $m_{b} = 4$ GeV, while the observed mass
difference is found to be $0.0487^{+0.0023}_{-0.0021}$ GeV~\cite{Tanabashi:2018oca}.
Using these values we find the spin averaged mass for $\bar ss$  as
\begin{equation}
m_{\bar ss} = 0.925 \ {\rm  GeV} 
\end{equation}
for $m_{\phi} = 1.019$ GeV. 
With the $\Lambda$ mass $m_{\Lambda} = 1.116$ GeV, 
the mass formula~\eqref{eq:mass-rel} suggests that a scalar meson $f_{0}$ composed of
two diquarks $\overline{ud} ud$ has a mass
\begin{equation}
   	m_{f_0} = 1.320\mathrm{\ GeV}.
\end{equation}
The corresponding particle can be found as $f_{0}(1370)$ 
in the particle listing of Particle Data Group~\cite{Tanabashi:2018oca},
in which the scalar resonance $f_{0}(1370)$ is reported as a broad resonance having a pole mass 
at $(1200\, \mathchar`-\, 1500) - i(150\, \mathchar`-\, 250)$ MeV. 
The mass of $f_{0}(1370)$ includes our value as a central value, and $f_{0}(1370)$ can be a
two-diquark state. For further confirmation, we need to investigate whether 
the property of $f_{0}(1370)$ shows the nature of a bound state of two diquarks. 
Such investigation could reveal the existence of the $ud$ diquark inside hadrons as an
effective degrees of freedom.

It is also interesting to mention that the mass differences of $m_{f_{0}} - m_{\Lambda}$ and
$m_{\Lambda} - m_{\bar ss}$ are about 200 MeV, which is the consequence of the V(3) symmetry breaking
in the nonet represetation. 
Similarly, the V(3) breaking appearing in the triplet representations shown in Sec.~\ref{sec: triplet rep} 
can be found in the mass differences of $m_{\Lambda_{c}}- m_{\bar sc}$ and 
$m_{\Lambda_{b}}-m_{\bar sb}$, where $m_{\bar s c}$ and $m_{\bar sb}$ stand for 
the spin averaged masses of $D_{s}^{*} \mathchar`- D_{s}$ and $B_{s}^{*} \mathchar`- B_{s}$, respectively, 
and these values are also about 200 MeV. If this V(3) breaking found in these hadrons is attributed to 
the mass difference between the $s$ quark and the $ud$ diquark, 
the mass of the $ud$ diquark may be 700 MeV if one assumes the constituent strange quark mass 
to be 500 MeV. 
Nevertheless, it should be worth mentioning that the V(3) breaking could stem from asymmetry 
of the interaction of $\bar s \mathchar`- Q$ and $(ud) \mathchar`- Q$ 
as pointed out in Refs.~\cite{Jido:2016yuv,Kumakawa:2017ffl}. 
There they found that the string tension in the color electric confinement potential 
between quark and diquark is as weak as half of that between quark and anti-quark,
even though these two systems have the same color configuration. 
Further investigation on the symmetry 
between the quark and diquark should be necessary.  

\subsection{ Quintet representation}

{The $\Psi\Psi c$ configuration has the quintet representation for the ground state without orbital excitations. In the symmetric limit, the $\Omega_{c}$ baryons with spin 1/2 and 3/2 and tetraquark mesons $\overline{ud}sc$ with spin 0 and 1 get degenerate. The symmetry breaking stems from the mass difference between the $s$ quark and  the $\overline{ud}$ diquark and the spin-spin interaction among the $s$ and $c$ quarks. The mass difference of the $s$ quark and the $\overline{ud}$ diquark is to be obtained as 200 MeV in the previous section. The mass splitting due to the spin-spin interaction can be removed by taking the spin averaged mass. The spin averaged mass of the $\Omega_{c}$ baryons is  observed as 2.742~GeV. Thus, we estimate the spin averaged mass of the tetraquarks $\overline{ud}sc$ to be 2.942~GeV.  }

{It is also interesting to estimate the mass of the tetraquark $\overline{ud}sb$ appearing the quintet representation of the $\Psi\Psi b$ configuration.  Only the $\Omega_{b}$ baryon with spin 1/2 is shown in the particle date table~\cite{Tanabashi:2018oca} and its mass is observed as 6.046~GeV. The mass splitting of the $\Omega_{b}$ barons with spin 1/2 and 3/2 due to the spin-spin interaction can be estimated as 0.023~GeV, because the spin-spin interaction is inversely proportional to the quark mass and the mass difference between the $\Omega_{c}$ baryons with spin 1/2 and 3/2 is observed as $0.0707^{+0.0008}_{-0.0009}$~GeV. Here we have assumed the charm and bottom quark masses as 1.3~GeV and 4.0~GeV, respectively. Therefore, the spin-averaged mass of the $\Omega_{b}$ baryons is estimated as 6.061~GeV, we find the spin-averaged mass of the tetraquark $\overline{ud}sb$ to be 6.261~GeV.  }


\section{Summary}
We have introduced a symmetry among the constituent strange quark and the $\overline{ud}$ diquark, 
supposing that their masses be very similar to each other, say 500 MeV.  
To investigate the properties of this symmetry, we have constructed an algebra which transforms a fermion with spin 1/2 and a boson with spin 0. Regarding these fermion and boson as a fundamental representation of this algebra, we have built higher representations for mesonic, diquark and baryonic configurations. We have proposed possible assignments of these irreducible representations to existing hadrons, which is summarized in Table~\ref{tab:multiplets}. Particularly investigating the triplet and nonet representations, we have found that ($\Lambda_{c}$, $D_{s}$, $D_{s}^{*}$) and ($\eta_{s}$, $\phi$, $\Lambda$, $f_{0}(1370)$) could form multiplets, respectively. Introducing a symmetry breaking coming from the mass difference the $s$ quark and the $\overline{ud}$ diquark, we have derived a mass relation among each multiplet. 
In the nonet representation, we have the mass relation among $\phi$, $\eta_s$, $\Lambda$, $f_0$. 
In our formulation, both $\phi$ and $\eta_s$ are composed of the $s$ and $\bar s$ quarks,
while the physical $\eta$ meson is known to be expressed almost as the octet. 
Thus, estimating the strength of the spin-spin interaction from the mass difference of $D_{s}$ and $D_{s}^{*}$, we have found the spin averaged mass $\phi$ and $\eta_{s}$ to be 920 MeV.  Using the mass relation with this mass and the observed $\Lambda$ mass, we have found the mass of $f_{0}$ in the multiplet to be 1320 MeV. This may correspond to the observed  $f_{0}(1370)$ meson. Thus, our model suggests $f_{0}(1370)$ to be a tetraquark composed of $ud$ and $\overline{ud}$ diquarks. In addition, finding the mass differences among the nonet to be 200 MeV,
and the difference between the spin averaged mass of $D_{s}$ and $D_{s}^{*}$ ($B_{s}$ and $B_{s}^{*}$)  and the $\Lambda_{c}$ ($\Lambda_{b}$) mass also to be 200 MeV, we have suggested the mass difference between the constituent $s$ quark and the $ud$ diquark to be 200 MeV. Thus, if we regard the strange quark mass as 500 MeV, the mass of $ud$ diquark may be 700 MeV. 
{For the $\Psi\Psi c$ configuration, we have found that the $\Omega_{c}$ baryons with spin 1/2 and 3/2 and tetraquark mesons $\overline{ud}sc$ with spin 0 and 1 are in the same multiplet. Estimating the mass difference between the baryons and mesons from the mass difference of the $s$ quark and $\overline{ud}$ diquark, we have found possible masses of the tetraquarks $\overline{ud}sc$ and $\overline{ud}sb$ to be 2.942~GeV and 6.261~GeV, respectively.}

{As a future study, it would be interesting if we could discuss the symmetry between the $s$ quark and the $\overline{ud}$ diquark in the hadron production, for instance, from $p\bar p$ collisions. Certainly we could have some symmetry or asymmetry in the productions of the $s$ quark and $\overline{ud}$ diquark. Such symmetry could be seen in the production rates of the hadrons. }
In particularly, because the $\Lambda$ hyperon with the strange quark has likewise a quark-diqurak structure as well as the $\Lambda_c$ baryon with the charm quark, their production mechanism would be quite similar. Such similarity could be seen in their production rate. 
{It should be also mentioning that} we could have another possibility for the source of the mass difference to be a perspective suggesting that the symmetry braking comes from the difference of interactions between ${\bar s} \mathchar`- Q$ and $(ud) \mathchar`- Q$. It is an open question that the origin of the symmetry breaking between the $s$ quark and the $\overline{ud}$ diquark, and further investigations on this issue are necessary.

\section*{Acknowledgment}
The work of D.J.\ was partly supported by Grants-in-Aid for Scientific Research from JSPS (17K05449).


\begin{thebibliography}{99}

\bibitem{Neeman:1961jhl} 
  Y.~Ne'eman,
  Nucl.\ Phys.\  {\bf 26}, 222 (1961).
  doi:10.1016/0029-5582(61)90134-1

\bibitem{GellMann:1962xb} 
  M.~Gell-Mann,
  Phys.\ Rev.\  {\bf 125}, 1067 (1962).
  doi:10.1103/PhysRev.125.1067
  
\bibitem{Nakano:1953zz} 
  T.~Nakano and K.~Nishijima,
  Prog.\ Theor.\ Phys.\  {\bf 10}, 581 (1953).
  doi:10.1143/PTP.10.581

\bibitem{GellMann:1953zza} 
  M.~Gell-Mann,
  Phys.\ Rev.\  {\bf 92}, 833 (1953).
  doi:10.1103/PhysRev.92.833

\bibitem{GellMann:1964nj} 
  M.~Gell-Mann,
  Phys.\ Lett.\  {\bf 8}, 214 (1964).
  doi:10.1016/S0031-9163(64)92001-3

\bibitem{Okubo:1961jc} 
  S.~Okubo,
  Prog.\ Theor.\ Phys.\  {\bf 27}, 949 (1962).
  doi:10.1143/PTP.27.949


\bibitem{Ida:1966ev}
  M.~Ida and R.~Kobayashi,
  Prog.\ Theor.\ Phys.\  {\bf 36}, 846 (1966).

\bibitem{Lichtenberg:1967zz}
  D.~B.~Lichtenberg and L.~J.~Tassie,
  Phys.\ Rev.\  {\bf 155}, 1601 (1967).

\bibitem{Anselmino:1992vg}
M.~Anselmino, E.~Predazzi, S.~Ekelin, S.~Fredriksson, and D.B.~Lichtenberg,
Rev.\ Mod.\ Phys.\ {\bf 65}, 1199 (1993).

\bibitem{Jaffe:2004ph}
  R.~L.~Jaffe,
  Phys.\ Rept.\  {\bf 409}, 1 (2005)
  [Nucl.\ Phys.\ Proc.\ Suppl.\  {\bf 142}, 343 (2005)]


\bibitem{Hess98}
  M.~Hess, F.~Karsch, E.~Laermann and I.~Wetzorke,
  Phys.\ Rev.\  D {\bf 58}, 111502 (1998).
 
 \bibitem{Babich}
  R.~Babich {\it et al.}
  J.\ High Energy Phys.\  {\bf 01} (2006) 086.

\bibitem{Orginos}
  K.~Orginos,
  Proc.\ Sci.,\ LAT2005 (2005)054.

\bibitem{Alexandrou}  
  C.~Alexandrou, P.~de Forcrand and B.~Lucini,
  Phys.\ Rev.\ Lett.\  {\bf 97}, 222002 (2006).

\bibitem{Goldstein:1979wba}
G.R.~Goldstein and J.~Maharana,
Nuovo Cim. {\bf A59}, 393 (1980).

\bibitem{Lichtenberg:1981pp}
D.B.~Lichtenberg,
Phys. Rev. {\bf 178}, 2197 (1969).


\bibitem{Lichtenberg:1982jp}
D.B.~Lichtenberg, W.~Namgung, E.~Predazzi, and J.G.~Wills,
Phys. Rev. Lett. {\bf 48}, 1653 (1982).

\bibitem{Liu:1983us}
K.F.~Liu and C.W.~Wong,
Phys. Rev. {\bf D28}, 170 (1983).

\bibitem{Ebert:2005xj}
D.~Ebert, R.N.~Faustov, and V.O.~Galkin,
Phys. Rev. {\bf D72} (2005) 034026.

\bibitem{Ebert:2007nw}
D.~Ebert, R.N.~Faustov, and V.O.~Galkin,
Phys. Lett. {\bf B659} (2008) 612.

\bibitem{Hernandez:2008ej}
E.~Hernandez, J.~Nieves, and J.M.~Verde-Velasco,
Phys.\ Lett.\ {\bf B666}, 150 (2008).

\bibitem{Lee:2009rt}
S.H. Lee and Shigehiro Yasui,
Eur.\ Phys.\ J.\ {\bf C64},  283 (2009).

\bibitem{Jido:2016yuv} 
  D.~Jido and M.~Sakashita,
  PTEP {\bf 2016}, 083D02 (2016)
  doi:10.1093/ptep/ptw113.

\bibitem{Kumakawa:2017ffl} 
  K.~Kumakawa and D.~Jido,
  PTEP {\bf 2017}, no. 12, 123D01 (2017)
  doi:10.1093/ptep/ptx155.
  
\bibitem{Kim:2011ut}
K.~Kim, D.~Jido, and S.H.~Lee,
Phys. Rev. C {\bf 84}, 025204 (2011).

\bibitem{miya66}
  H.~Miyazawa,
  Prog.\ Theor.\ Phys.\  {\bf 36} (1966) no.6,  1266.
  doi:10.1143/PTP.36.1266

\bibitem{miya68}
  H.~Miyazawa,
  Phys.\ Rev.\  {\bf 170} (1968) 1586.
  doi:10.1103/PhysRev.170.1586

\bibitem{Catto:1984wi} 
  S.~Catto and F.~Gursey,
  Nuovo Cim.\ A {\bf 86}, 201 (1985).
  doi:10.1007/BF02902548
  
\bibitem{Lichtenberg:1989ix} 
  D.~B.~Lichtenberg,
  J.\ Phys.\ G {\bf 16}, 1599 (1990).
  doi:10.1088/0954-3899/16/11/010
  
\bibitem{Nielsen:2018uyn}
  M.~Nielsen and S.~J.~Brodsky,
  Phys.\ Rev.\ D {\bf 97} (2018) no.11,  114001
  doi:10.1103/PhysRevD.97.114001.
  
\bibitem{Iachello:1980av} 
  F.~Iachello,
  Phys.\ Rev.\ Lett.\  {\bf 44}, 772 (1980).
  doi:10.1103/PhysRevLett.44.772
  
\bibitem{Tanabashi:2018oca} 
  M.~Tanabashi {\it et al.} [Particle Data Group],
  Phys.\ Rev.\ D {\bf 98}, no. 3, 030001 (2018).
  doi:10.1103/PhysRevD.98.030001


\end{thebibliography}
\end{document}